\documentclass[a4paper,12pt]{article}

\usepackage{amssymb,amsmath}
\usepackage[totalwidth=17cm,totalheight=24cm]{geometry}
\usepackage{graphicx}

\newcommand{\C}{{\mathbb C}}
\newcommand{\R}{{\mathbb R}}
\newcommand{\Z}{{\mathbb Z}}

\newcommand{\cM}{{\cal M}}
\newcommand{\bK}{{\bar{K}}}
\newcommand{\bP}{{\bar{P}}}
\newcommand{\id}{{\mathbb I}}
\newcommand{\im}{{\rm i\,}}

\newcommand{\x}{{\tilde{x}}}

\newcommand{\dirac}{{\backslash\!\!\!\partial}}

\newcommand{\be}{\begin{eqnarray}}
\newcommand{\ee}{\end{eqnarray}}

\begin{document}
 \pagestyle{plain}
\title{Fermions,  differential forms and doubled geometry}
\author{ Kirill Krasnov \\ {}\\
{\small \it School of Mathematical Sciences, University of Nottingham, NG7 2RD,
UK}}

\date{March 2018}
\maketitle

\begin{abstract}\noindent We show that all fermions of one generation of the Standard Model (SM) can be elegantly described by a single fixed parity (say even) inhomogeneous real-valued differential form in seven dimensions. In this formalism the full kinetic term of the SM fermionic Lagrangian is reproduced as the appropriate dimensional reduction of $(\Psi, D\Psi)$ where $\Psi$ is a general even degree differential form in $\R^7$, the inner product $(\cdot,\cdot)$ is as described in the main text, and $D$ is essentially an appropriately interpreted exterior derivative operator. The new formalism is based on geometric constructions originating in the subjects of generalised geometry and double field theory. 
\end{abstract}

\section{Introduction}

The gauge group of the Pati-Salam unification model \cite{Pati:1973uk} is ${\rm SO}(4)\sim {\rm SU}(2)\times{\rm SU}(2)$ times ${\rm SO}(6)\sim {\rm SU}(4)$. This last group extends the usual ${\rm SU}(3)$ of the Standard Model (SM) by interpreting the lepton charge as the fourth colour. The groups  ${\rm SO}(4)$ and ${\rm SO}(6)$ are usually put together into $ {\rm SO}(4)\times{\rm SO}(6)\subset{\rm SO}(10)$ unified gauge group, see e.g. \cite{Baez:2009dj} for a nice exposition of this standard material. 

It has long been known that the Pati-Salam group ${\rm SO}(4)\times{\rm SO}(6)$ can be put together with the Lorentz group ${\rm SO}(3,1)$ into a pseudo-orthogonal group in dimension 14 so that one generation of the Standard Model fermions is described by a single spinor representation of this large gauge group. In particular, a model based on ${\rm SO}(3,1)\times{\rm SO}(10)\subset {\rm SO}(3,11)$ has been studied in the literature, see \cite{Nesti:2009kk} and also \cite{Krasnov:2017epi} for a recent review of the related material. This group has signature zero (modulo eight), and so its Weyl representations are real. It can then be seen that the fermions of a single generation of the SM, including their Lorentz spinor components, can be described by a single real Weyl representation of ${\rm SO}(3,11)$. Further, the Weyl Lagrangian in $\R^{3,11}$ dimensionally reduces to just the right collection of chiral Dirac Lagrangians for the SM elementary particles in four dimensions. 

It has also been previously noted, in particular in \cite{Maraner:2003sq}, that there is another group for which a similar construction is possible. Namely, given that the non-compact Lorentz group ${\rm SO}(3,1)$ is part of the game, there is no reason to restrict one's attention to only compact gauge groups on the SM side. Thus, we can also embed ${\rm SO}(4)\times{\rm SO}(6)\subset{\rm SO}(4,6)$, and the latter fits together with the Lorentz group into ${\rm SO}(3,1)\times{\rm SO}(4,6)\subset{\rm SO}(7,7)$. The representation theory in this case works out correctly as well, and it can similarly be expected that the Weyl Lagrangian in $\R^{7,7}$ dimensionally reduces to the right collection of chiral Dirac Lagrangians in four dimensions. 

The split (or neutral) signature groups like ${\rm SO}(7,7)$ also appear and play central role in completely different subjects of generalised geometry \cite{Hitchin:2004ut}, \cite{Hitchin:2010qz} and double field theory \cite{Hull:2009mi}. In particular, in the generalised geometry setup it is of central importance that spinors of ${\rm SO}(n,n)$ have the natural interpretation of differential forms in $\R^n$. 

The purpose of this article is to combine the referred to above representation theory construction putting together the Lorentz group ${\rm SO}(3,1)$ with the Pati-Salam group ${\rm SO}(4)\times{\rm SO}(6)$ into ${\rm SO}(7,7)$ with the geometric setups of generalised geometry and double field theory. What is achieved as the result is an interpretation of the SM fermions as even degree differential forms in seven dimensions. Further, we achieve a very compact rewriting of the kinetic part of the SM fermionic Lagrangian. Thus, the main result of this article can be summarised by the following relation
\be\label{main-relation}
(\Psi,D\Psi) =  \frac{\im}{2} (\xi^\dagger)^{aI} \dirac \xi_{aI} - \frac{\im}{2} (\bar{\xi}^\dagger)^{aI} \dirac \bar{\xi}_{aI} + {\rm c.c.}
\ee
Here $\Psi\in \Lambda^{even} \R^7$ is a general inhomogeneous even degree real form in $\R^7$, the ${\rm SO}(7,7)$ invariant inner product $(\cdot,\cdot)$ is to be explained below, but is essentially a combination of an involution on $\Lambda^\bullet \R^7$ and restriction to the top form, and $D$ is the Dirac operator on $\R^{7,7}$ that is essentially an appropriately interpreted exterior derivative operator, see below. All differential form component functions are originally assumed to be functions on $\R^{7,7}$, the double of the space on which the differential forms are taken. This makes the subject of double field theory with some of its geometric constructions relevant. 

To achieve a reduction to the right-hand-side in (\ref{main-relation}) one assumes that the component functions only depend on 4 of the coordinates on $\R^{7,7}$, so that they are in fact functions on a copy of Minkowski space $\R^{3,1}$. This involves a choice of which copy of $\R^{3,1}$ is taken, and it is in this process of selecting ${\rm SO}(3,1)\subset{\rm SO}(7,7)$ that the metric to which all fermions on the right-hand-side of (\ref{main-relation}) couple arises. Then $\dirac$ is the chiral Dirac operator that maps unprimed 2-component Lorentz spinors to primed 2-component spinors, and $\xi_{aI}, \bar{\xi}_{aI}$ are all unprimed 2-component Lorentz spinors. The indices $a=1,2$ are the isospin ones, and $I=1,2,3,4$ are the colour ones, with the lepton charge interpreted as the fourth colour. The spinors $\xi_{aI}$ are spinor representations of the ${\rm SU}(2)_L$, and thus describe left-handed particles, while $\bar{\xi}_{aI}$ are representations of the ${\rm SU}(2)_R$. If desired, the right-hand-side in (\ref{main-relation}) can be further split by choosing ${\rm SU}(3)\subset {\rm SU}(4)$, and then rewriting everything in terms of the usual leptons and quarks. We explain everything in more details in the main text. 

Thus, the main outcome of this article is a geometric construction which makes it obvious that SM fermions are, or at least can be elegantly described by, differential forms. 

There have been numerous previous attempts to interpret spinors as differential forms. Indeed, it has been known for a very long time that spinors are related to differential forms, and the Dirac operator is related to the exterior derivative operator. In the physics literature this has been studied under the name of Dirac-K\"ahler fermions, the approach initiated in \cite{Kahler}. The other well-known references on this approach are \cite{Graf:1978kr}, \cite{Benn:1982sr}. 

Briefly, the idea is to consider the first-order differential operator $d+d^*$, where $d$ is the exterior derivative and $d^*$ is the dual operator. This operator is the square root of the Laplacian operator $d d^* + d^* d$ on differential forms, and naturally acts on the space of inhomogeneous differential forms (the space of differential forms of all degrees). Given that there is a relation between the Clifford algebra over a vector space and the exterior algebra, see below for a review of this, the space of differential forms is a module for the Clifford algebra, and thus has a spinor interpretation. 

There is, however, the following well-known problem with this idea. After differential forms are given spinor interpretation they turn out to carry integer spinor representations. This can be seen in many different ways. A particularly straightforward way is available in four dimensions and uses the 2-component spinor formalism. Indeed, 1- and 3-forms are objects $v_\mu$ with one spacetime index, which translates into two spinor indices of different type $v_\mu \to v_{MM'}$. The 2-forms $v_{\mu\nu}$ can be decomposed into their self- and anti-self-dual parts, and these correspond to rank 2 spinors $v_{MN}, v_{M'N'}$. All in all, differential forms are bi-spinors rather than spinors. In the cited above literature this fact is expressed by saying that in four dimensions a Dirac-K\"ahler fermion is a collection of four Dirac fermions. 

While the above mentioned problem makes the old ideas \cite{Kahler} of little use in physics, the relation between differential forms and spinors has found a much more respectful place in mathematics, where it is regarded as classical. Thus, as is well-known, the Clifford algebra over a vector space $V$ with inner product and the exterior algebra over $V$ are isomorphic as vector spaces. Moreover, and this fact will play the central role in the present article, the spinor representations of the orthogonal group ${\rm SO}(2n)$ can be realised in the space $\Lambda^\bullet \R^n$ of inhomogeneous differential forms {\it in half the dimension}. In the physics literature this construction goes under the name of (fermionic) oscillator realisation of representations. Another classical geometrical construction that makes prominent use of differential forms and the natural Dirac-like operator $D=d+d^*$ is the article \cite{Atiyah} that used the index theorem for $D$ to prove properties of zeros of vector fields on a manifold. 

In this article we combine the mentioned construction of spinor representations of ${\rm SO}(2n)$ in the space $\Lambda^\bullet \R^n$ with the geometric setup of double field theory \cite{Hull:2009mi}. This leads to a realisation of spinors as differential forms, with the Dirac operator being related to the exterior derivative operator. In the construction that we describe the problem of the Dirac-K\"ahler approach that differential forms are bi-spinors rather than spinors does not arise. This is achieved by working with differential forms in a space of half the dimension, which explains why the geometric setup of double field theory is particularly relevant. 

The main points of our construction are quite simple and can be explained already in the Introduction. Let us start by describing the referred to differential forms construction of spinor representations of orthogonal groups. We will restrict our attention to the split  signature orthogonal groups ${\rm SO}(n,n)$ for which this construction is particularly elegant. Thus, we start with a vector space $V$ of dimension $2n$ that is equipped with a metric of split (neutral) signature. For any such space we can choose a doubly-null basis, i.e. 
\be\label{V-split}
V= T\oplus \tilde{T},
\ee
so that both $T,\tilde{T}$ are totally null, see more on the geometry of such a choice below. We will represent elements of this vector space as columns 
\be\label{vec-V}
V\ni X = \left( \begin{array}{c} \xi \\ \eta \end{array} \right), \qquad \xi\in T, \eta\in \tilde{T},
\ee
The split signature metric is 
\be\label{G}
G( (\xi_1,\eta_1),(\xi_2,\eta_2)) := G(\xi_1,\eta_2) + G(\eta_1,\xi_2),
\ee
where $G(\xi,\eta)$ is some (non-degenerate) pairing that provides an identification $\tilde{T}\sim T^*$ of objects $\eta$ with duals of objects $\xi$. This geometric setup is present in double field theory \cite{Hull:2009mi}, and also in a related but different way in Hitchin's generalised geometry \cite{Hitchin:2004ut}, \cite{Hitchin:2010qz}.\footnote{In particular, in the generalised geometry setup the vector space $V$ is the direct sum of spaces of vectors and 1-forms $V=TM\oplus T^*M$, first taken at a given point of a manifold $M$, and the split signature metric is given by $G( (\xi_1,\eta_1),(\xi_2,\eta_2)) = (1/2)\left(\eta_1(\xi_2) + \eta_2(\xi_1)\right)$.}

Given a vector space equipped with an inner product, we form the Clifford algebra. Thus, we define the Clifford algebra for $V$ as the algebra generated by $V$ modulo the defining relation 
\be\label{clifford}
X_1X_2 + X_2X_1 = 2 G(X_1,X_2) \id, \qquad X_{1,2}\in V.
\ee
This Clifford algebra has a natural representation in the exterior algebra $\Lambda^\bullet \tilde{T}$ over $\tilde{T}$. Indeed, the Clifford multiplication of $\xi+\eta\in V$ with a general element $\phi\in \Lambda^\bullet \tilde{T}$ is defined as follows
\be\label{action-forms}
c(\xi+\eta) \phi := i_\xi \phi + \eta\wedge \phi,
\ee
where the interior product $i_\xi$ is defined on elements of $\tilde{T}$ via $i_\xi \eta :=2 G(\xi,\eta)$, and this is extended to arbitrary forms by linearity. This defines a representation because
\be
c(\xi_1+\eta_1)c(\xi_2+\eta_2)\phi = i_{\xi_1} ( i_{\xi_2}\phi + \eta_2\wedge \phi) + \eta_1\wedge  ( i_{\xi_2}\phi + \eta_2\wedge \phi) \\ \nonumber =
i_{\xi_1\xi_2} \phi + (i_{\xi_1}\eta_2)\phi - \eta_2\wedge i_{\xi_1}\phi + \eta_1\wedge i_{\xi_2}\phi + \eta_1\wedge\eta_2\wedge\phi,
\ee
which shows that
\be
c(X_1 X_2 + X_2 X_1)\phi = (i_{\xi_1}\eta_2 + i_{\xi_2}\eta_1)\phi = 2G(X_1,X_2)\phi,
\ee
and so we have (\ref{clifford}). This construction identifies spinors of ${\rm SO}(T\oplus \tilde{T})$ with elements of $\Lambda^\bullet \tilde{T}$. In particular, Weyl spinors are identified with the spaces of even/odd elements in $\Lambda^\bullet \tilde{T}$. In the generalised geometry approach \cite{Hitchin:2004ut} this construction gives identification of spinors with differential forms, and plays the central role. 

We now extend the above linear algebra construction into a differential geometry one. For this we use the geometric setup of double field theory. It is at this point where we start to deviate from the generalised geometry setup. As is explained in \cite{Vaisman:2012ke}, see also \cite{Freidel:2017yuv}, the geometry of double field theory is that of a flat bi-Lagrangian, or para-K\"ahler manifold. Thus, let $\cM$ be a manifold of dimension $2n$ endowed with a split (neutral) signature metric $G$ and a metric-compatible para-complex structure $F$ such that 
\be\label{F-compat}
F^2=\id \quad {\rm and}\quad  G(FX,FY)=-G(X,Y).
\ee 
The para-complex structure $F$ splits $T \cM = T\oplus \tilde{T}$ the tangent space $T \cM$ into eigenspaces $T, \tilde{T}$ of eigenvalue $\pm 1$. The minus sign in the metric compatibility relation implies that both these eigenspaces are null. "Lowering the index" on $F$ with $G$ we get another tensor $W(X,Y):=G(FX, Y)$ that is anti-symmetric $W(Y,X)=-W(X,Y)$. Then the fact that $T, \tilde{T}$ are null implies that the distributions $T, \tilde{T}$ are Lagrangian with respect to the symplectic form $W$. All this statements will be verified in the main text. For simplicity, we shall assume throughout this article that the metric $G$ is flat and distributions $T,\tilde{T}$ are integrable. 

It should be noted that the described double field theory setup is a real version of the usual setup of complex manifolds with their compatible triple of a symplectic form, a Hermitian metric, a complex structure and an integrable distribution of the tangent space into $(1,0)$ and $(0,1)$ subspaces. Moreover, in the complex manifold setup it is well-known that the Dirac operator is essentially the Dolbeault operator $\bar{\partial}$. One has $D = \bar{\partial}+\bar{\partial}^* : \Lambda^{0,{\rm even}} \to \Lambda^{0,{\rm odd}}$, see e.g. Exercise 2.3.39 in \cite{Freed}. Thus, in the complex manifolds case the spinors are differential forms in half the dimension, i.e. those generated by elementary forms $d\bar{z}^i$, where $z^i,\bar{z}^i$ are the complex coordinates.  

What we will describe next can be interpreted as the real version of the construction of the previous paragraph.  Let $\cM$ be a para-K\"ahler manifold with a split signature metric $G$ and a compatible doubly-null integrable distribution $T\cM = T \oplus \tilde{T}$. Let $T^*\cM=T^* \oplus \tilde{T}^*$ be the corresponding distribution of the cotangent space. Let $x^I$ be a set of coordinates for $\cM$, with $x^I=(x^i, \tilde{x}_i), i=1,\ldots, n$ and the corresponding basis of $T^* \cM$ being given by 1-forms
\be\label{basic-1-forms}
dx^i \in T^*, \qquad d\tilde{x}_i \in \tilde{T}^*,
\ee
so that the metric pairing is given by
\be
G(dx^i , d\tilde{x}_j) = \frac{1}{2} \delta^i_j.
\ee
We now form the Clifford algebra for $T^* \oplus \tilde{T}^*$. It is generated by $dx^i, d\tilde{x}_i$ with the defining relations being
\be\label{dx-clifford}
dx^i d\tilde{x}_j + d\tilde{x}_j dx^i = \delta^i_j,
\ee
and both $dx^i, d\tilde{x}_j$ anti-commuting between themselves. As we already know, the space of spinors for ${\rm SO}(n,n)$ is naturally identified with the space $\Lambda^\bullet T^*$ whose elements are differential forms
\be\label{phi-intr}
\Lambda^\bullet T^* \ni \phi = \sum_{k=0}^n \frac{1}{k!} \phi_{i_1\ldots i_k}(x,\tilde{x}) dx^{i_1} \ldots dx^{i_k},
\ee
where the wedge product is implied. The Dirac operator for the metric $G$ is then explicitly described as follows
\be
D = c( dx^I) \frac{\partial}{\partial x^I} = c( dx^i) \frac{\partial}{\partial x^i} + c(d\tilde{x}_i) \frac{\partial}{\partial \tilde{x}_i},
\ee
where $c$ is the Clifford multiplication. Then, as we know from (\ref{action-forms}), Clifford multiplication $c(dx^i)$ is just the wedge product, while $c(d\tilde{x}_i)$ acts by interior multiplication. Explicitly,
\be\label{dirac}
D \phi = \sum_{k=0}^n \frac{1}{k!}  \partial_i \phi_{i_1\ldots i_k}(x,\tilde{x}) dx^i dx^{i_1} \ldots dx^{i_k} \\ \nonumber +\sum_{k=0}^n \frac{1}{k!} 
\tilde{\partial}^i \phi_{i_1\ldots i_k}(x,\tilde{x}) \left( \delta_i^{i_1} dx^{i_2} \ldots dx^{i_k} + \ldots (-1)^{k-1} dx^{i_1} \ldots dx^{i_{k-1}} \delta_i^{i_k}\right),
\ee
where $\partial_i := \partial/\partial x^i, \tilde{\partial}^i := \partial/\partial \tilde{x}_i$. Thus, we see that the Dirac operator on $\cM$ for a ${\rm SO}(n,n)$ metric $G$ is just the appropriately interpreted exterior derivative operator. Numerous explicit examples of working with the operator $D$ will be given in the main text. We hope that the reader will appreciate the naturalness of this construction, in particular by following these examples. 

What we have described is just the translation of well-known construction of the Dirac operator on K\"ahler manifolds to the real setup of para-K\"ahler manifolds, but, as far as we know, this translation has not appeared in the literature before. This is presumably due to the fact that the construction of the spinor representation of ${\rm SO}(n,n)$ as that in the space $\Lambda^\bullet \R^n$ is most frequently met in the geometric setup of generalised geometry, where the manifold remains of dimension $n$ and in particular not doubled, while we saw this to be necessary to describe the Dirac operator. On the other hand, in the double field theory framework spinors and differential forms has not  played any significant role up to now, and so there was no motivation to consider the Dirac operator. 

We can now explain how the other part of the geometric setup of double field theory, namely another metric on $\cM$ comes into play. This happens in the process of selecting of which ${\rm SO}(3,1)\subset{\rm SO}(7,7)$ is identified with the Lorentz group. Geometrically, this is done by selecting another decomposition $V=U\oplus \tilde{U}$ so that the metric $G$ restricts to a non-degenerate metric on $U,\tilde{U}$, e.g. positive definite on $U$ and negative definite on $\tilde{U}$. One can then identify the Lorentz group as that of mixing say first 3 of the directions in $U$ and one direction in $\tilde{U}$. Then, as we explain in the main text, the required decomposition $V=U\oplus \tilde{U}$ is the same as a metric $G$ compatible endomorphism $J$ squaring to identity $J^2=\id$, with $U,\tilde{U}$ being its eigenspaces. As we shall see in the main text, this endomorphism is essentially the generalised metric that arises in both double field theory and generalised geometry. Thus, this second geometric ingredient of double field theory also plays an important role in our construction, and arises in identifying the Lorentz group ${\rm SO}(3,1)$ inside ${\rm SO}(7,7)$.

The organisation of the rest of this paper is as follows. In Section \ref{sec:general}, we start  by spelling out the sketched in the Introduction geometric constructions in more details. We then study groups ${\rm SO}(n,n)$ of increasing dimension and work out what the described idea of embedding the Lorentz group ${\rm SO}(3,1)$ into ${\rm SO}(n,n)$ gives in each case. We will see that the properties of the ${\rm SO}(n,n)$ invariant inner product $(\cdot, \cdot)$ needed on the left-hand-side of (\ref{main-relation}) are such that only in very few cases the Lagrangian $(\Psi,D\Psi)$ produces something non-trivial. One of these cases is the setup of ${\rm SO}(7,7)$ that is related to the Standard Model. 

We start in Section \ref{sec:22} with the setup of ${\rm SO}(2,2)$. There is no Lorentz group inside in this case, and we just work out the spinor representations and explicitly verify that (\ref{dirac}) is the correct Dirac operator in this case. Our next example is that of ${\rm SO}(3,3)$, which we treat in Section \ref{sec:33}. The group ${\rm SO}(3,3)$ contains the Lorenz group. The subgroup that commutes with Lorenz group is ${\rm SO}(2)$, and so we expect to see charged fermions in this case. We verify that the Lagrangian $(\Psi,D\Psi)$ in this case reduces to the Weyl Lagrangian for a single charged 2-component Weyl fermion in four dimensions. As we explicitly verify in Section \ref{sec:55}, the Lagrangian $(\Psi,D\Psi)$ vanishes in the setup of ${\rm SO}(5,5)$. The setup of ${\rm SO}(7,7)$ is treated in Section \ref{sec:77}, where we explicitly verify that the SM fermion content arises and check the relation (\ref{main-relation}). We conclude with a discussion. 

\section{Geometric preliminaries}
\label{sec:general}

\subsection{Pseudo-orthogonal group ${\rm SO}(n,n)$}

The group of transformations preserving the metric (\ref{G}) is ${\rm O}(n,n)$. In this paper we are not interested in subtleties related to discrete subgroups, and so we will just denote the relevant group by ${\rm SO}(n,n)$ everywhere. Its Lie algebra can be described explicitly as follows. We follow \cite{Gualtieri:2003dx} closely, making necessary changes to work in the double field theory rather than generalised geometry setup. The Lie algebra of the group ${\rm SO}(n,n)$ consists of endomorphisms of $V$ with the property
\be
{\mathfrak so}(V) = \{ M | G(MX,Y)+ G(X,MY)=0\}.
\ee
To provide the explicit matrix description it is convenient to use the identification 
\be\label{T-ident}
\tilde{T}=T^*
\ee
that is provided by the metric $G$. With this identification in mind and the representation (\ref{vec-V}) of vectors in $V$ assumed, the Lie algebra ${\mathfrak so}(T\oplus T^*)$ consists of the following matrices
\be\label{M}
M = \left( \begin{array}{cc} A & \beta \\ B & -A^T \end{array}\right), \qquad A\in {\rm End}(T), \beta\in T\otimes T, B\in T^*\otimes T^*,
\ee
and both $\beta, B$ are anti-symmetric tensors. 

Explicitly, introducing a basis $e_i \in T$ and $\tilde{e}^i\in \tilde{T}$ so that the metric pairing is $G(e_i, \tilde{e}^j)=(1/2)\delta_i^j$, the general element of $V$ is $\xi^i e_i + \eta_i \tilde{e}^i$. The pairing of $T,\tilde{T}$ that is used in the identification (\ref{T-ident}) is $(\xi^i e_i, \eta_i \tilde{e}^i):= 2G(\xi^i e_i, \eta_i \tilde{e}^i) = \xi^i \eta_i$. The matrix $A$ the represents an endomorphism $\xi^i \to A^i{}_j \xi^j$. The tensors $\beta, B$ are objects $\beta^{ij}, B_{ij}$, and $A^T$ is the endomorphism $\eta_i \to A^j{}_i \eta_j$. A proof of the fact that $M$ preserves the metric $G$ is straightforward verification. 

The exponentiation of some of the subgroups is easy. 

\bigskip
\noindent {\bf B-transform.} Exponentiating the subgroup generated by $B$ we get
\be
\exp (B) = \left( \begin{array}{cc} 1 & 0 \\ B & 1 \end{array}\right). 
\ee
This acts on $X=(\xi,\eta)^T$ as
\be\label{B-action}
\exp(B) \circ X = \xi + \eta - i_\xi B.
\ee
The minus sign in the above formula is different from that in \cite{Gualtieri:2003dx}, but is more natural if the action of a 2-form on a vector field is in components $B_{ij} \xi^j$, i.e. the second index of the tensor in $O$ is contracted with the index in $X$. This convention also agrees more naturally with what we will see in the Clifford algebra. 

\bigskip
\noindent {\bf $\beta$-transform.} Exponentiating the subgroup generated by $\beta$ we get
\be
\exp(\beta) = \left( \begin{array}{cc} 1 & \beta \\ 0 & 1 \end{array}\right). 
\ee
This acts on $X$ as
\be\label{beta-action}
\exp(\beta) \circ X = \xi - i_\eta \beta + \eta .
\ee
Our sign here is also different from that in \cite{Gualtieri:2003dx}.

\bigskip
\noindent {\bf ${\rm GL}(n)$-transform.} Exponentiating the subgroup generated by $A$ we get
\be
\exp(A) = \left( \begin{array}{cc} \exp A & 0 \\ 0 & (\exp A^T)^{-1} \end{array}\right). 
\ee

\subsection{Spinors and differential forms}

As explained in the Introduction, we construct the Clifford algebra for $T\oplus T^*$. It is generated by the basis $e_i, \tilde{e}^i, e_i\in T, \tilde{e}^i \in T^*$ with the defining relations being
\be\label{e-clifford}
e_i \tilde{e}^j + \tilde{e}^j e_i = \delta_i^j,
\ee
and both $e_i, \tilde{e}^i$ mutually anti-commuting. The space $\Lambda^\bullet T^*$ is then a module for the above Clifford algebra with the Clifford multiplication being
\be\label{action-1}
c(e_i)  \tilde{e}^{i_1}  \ldots \tilde{e}^{i_k} = \delta_i^{i_1} \tilde{e}^{i_2} \ldots  \tilde{e}^{i_k} + \ldots + (-1)^{k-1} \tilde{e}^{i_1}  \ldots \tilde{e}^{i_{k-1}} \delta_i^{i_k},
\ee
and
\be\label{action-2}
c(\tilde{e}^i) \tilde{e}^{i_1} \ldots \tilde{e}^{i_k} = \tilde{e}^i  \tilde{e}^{i_1} \tilde{e}^{i_k} .
\ee
All this can also be described in a more physics-friendly creation-annihilation operator notation, see below.

\subsection{The action of ${\rm SO}(n,n)$ on spinors}

Our aim is to describe the action of subgroups of ${\rm SO}(n,n)$ on spinors as differential forms. Again, we follow \cite{Gualtieri:2003dx}. The group ${\rm SO}(T\oplus T^*)$ is doubly covered by ${\rm Spin}(T\oplus T^*)$, and the latter can be explicitly described as sitting inside the Clifford algebra 
\be\label{spin-group}
{\rm Spin}(T\oplus T^*) =\{ v_1\ldots v_r | v_i \in T\oplus T^*, G(v_i,v_i)=\pm 1\quad\! {\rm and}\quad\! r\quad\! {\rm even}\}.
\ee
The Lie algebra ${\mathfrak so}(V)$ is $\Lambda^2 V$, and this also sits naturally inside the Clifford algebra. Its action on $V$ can then be described as the natural action of $\Lambda^2 V$ on $V$ by the commutator, both viewed as sitting inside the Clifford algebra
\be
\omega\circ v = \omega v - v\omega, \qquad v\in V, \omega \in \Lambda^2 V.
\ee

Let us work this out in our setup. Using the Clifford algebra relations (\ref{e-clifford}) we have
\be
\tilde{e}^i \tilde{e}^j e_k - e_k \tilde{e}^i \tilde{e}^j  = 2 \tilde{e}^{[i} \delta_k^{j]},
\ee
and so for some 2-form $B=(1/2) B_{ij} \tilde{e}^i \tilde{e}^j$ its action on vectors is
\be
 [ \frac{1}{2} B_{ij} \tilde{e}^i \tilde{e}^j, \xi^k e_k] = e^i B_{ij} \xi^j = - i_\xi B.
\ee 
This is the same action as we have seen in (\ref{B-action}). 

Let us also see how the bi-vectors act. We have
\be
e_i e_j \tilde{e}^k - \tilde{e}^k e_i e_j = 2 e_{[i} \delta_{j]}^k,
\ee
and so
\be
 [\frac{1}{2} \beta^{ij} e_i e_j, \eta_k \tilde{e}^k] = e_i \beta^{ij} \eta_j = -i_\eta \beta,
\ee
which is the action from (\ref{beta-action}).

Finally, let us check that the action of the Lie algebra of ${\rm GL}(n)$ is also as we previously described. It corresponds to the commutator of the Clifford algebra element $(1/2)A^i{}_j (e_i \tilde{e}^j - \tilde{e}^j e_i)$ with $X$. Indeed, 
\be
 [\frac{1}{2} A^i{}_j (e_i \tilde{e}^j - \tilde{e}^j e_i) , \xi^k e_k + \eta_k \tilde{e}^k] = e_i A^i{}_j \xi^j - \eta_i A^i{}_j \tilde{e}^j,
\ee
which is the correct action. 

This allows us to write the action of the Lie algebra ${\mathfrak so}(T\oplus T^*)$ on spinors from $\Lambda^\bullet T^*$. We have
\be\label{action-M-forms}
c(M) \phi = c\left( \frac{1}{2} B_{ij} \tilde{e}^i \tilde{e}^j + \frac{1}{2} \beta^{ij} e_i e_j + \frac{1}{2} A^i{}_j(e_i \tilde{e}^j - \tilde{e}^j e_i) \right)  \phi.
\ee 
Using (\ref{action-1}) and (\ref{action-2}) this works out to 
\be
c(M)\phi = B\wedge \phi - i_\beta \phi - A^T\phi + \frac{1}{2} {\rm Tr}(A) \phi,
\ee
where $A^T\phi$ is the natural action of ${\rm GL}(n)$ on forms $(A^T\phi)_{i_1\ldots i_k} = k A^j{}_{[i_1} \phi_{|j| i_2 \ldots i_k]}$ and
\be
i_\beta \phi = \frac{1}{2 (k-2)!} \beta^{ij} \phi_{ij i_1 \ldots i_{k-2}} e^{i_1}\wedge e^{i_{k-2}}
\ee
is the insertion of the bi-vector $\beta$ into the $k$-form $\phi$. 

\subsection{Creating-annihilation operators}

The above Clifford algebra relations and the formula (\ref{action-M-forms}) can be rewritten in more physics friendly notations with the help of creation-annihilation operators. Thus, we identify
\be
(a^i)^\dagger:=\tilde{e}^i , \qquad a_i := e_i.
\ee
We then have
\be
a_i (a^j)^\dagger + (a^j)^\dagger a_i = \delta^j_i.
\ee
The Clifford algebra module $\Lambda^\bullet T^*$ is then the Hilbert space spanned by all vectors created from the vacuum $|\Omega \rangle$ via the creation operators
\be
(a^{i_1})^\dagger \ldots (a^{i_k})^\dagger |\Omega \rangle.
\ee
The formula (\ref{action-M-forms}) takes the following form
\be\label{operator-M}
M =  \frac{1}{2} B_{ij} (a^i)^\dagger (a^j)^\dagger + \frac{1}{2} \beta^{ij} a_i a_j + \frac{1}{2} A^i{}_j(a_i (a^j)^\dagger - (a^j)^\dagger a_i) .
\ee

\subsection{The double field theory setup and the Dirac operator}

In preparation for the description of the Dirac operator in the above language, we now describe the geometric setup of double field theory in some more detail. As already described in the Introduction, we start with a manifold $\cM$ of dimension $2n$ with a split signature metric $G$ on it. We then require that there exists a endomorphism of the tangent bundle $F: F^2=\id$, which is metric compatible in the sense of (\ref{F-compat}). This splits the tangent bundle into subspaces $T, \tilde{T}$ of eigenvalues $\pm 1$ of $F$. It is not hard to see that these subspaces are totally null. Indeed, denoting by $X', Y'$ eigenvectors of $F$ we have
\be
G(X', Y') = -G(FX' , FY') = - G(X', Y'),
\ee
and so $G(X', Y')=0$. The data of $F$ and $G$ define another tensor $W(X,Y)=G(FX,Y)$, which can be thought of as the endomorphism $F$ with one of its indices lowered with the metric $G$. This tensor is in this case anti-symmetric
\be
W(Y,X) = G(FY,X)= - G(FFY, FX) = - G(Y, FX) = -W(X,Y),
\ee
and so is a 2-form. It is easy to see that the subspaces $T,\tilde{T}$ are Lagrangian with respect to $W$. So, manifolds of this type can be referred to as bi-Lagrangian. 

In what follows we assume that the metric $G$ is flat, and that we can work in coordinates $x^i, \tilde{x}_i$ in which the basis in $T^*\cM$ is given by (\ref{basic-1-forms}). We then generate the Clifford algebra as in (\ref{dx-clifford}), and realise its spinor representations by elements in $\Lambda^\bullet T^*$, and so by differential forms of the type (\ref{phi-intr}). Then the Dirac operator for the metric $G$ is given by (\ref{dirac}). The formula (\ref{operator-M}) for the action of the Lie algebra of Lorentz group on forms still applies, one just has to identify
\be
(a^i)^\dagger = dx^i, \qquad a_i = d\tilde{x}_i,
\ee
and keep in mind that the fist operator acts by the usual wedge product, while the second operator acts by interior product
\be
c(d\tilde{x}_i) dx^j = \delta_i^j.
\ee

\subsection{The inner product(s)}

We now describe two different inner products in the space $\Lambda^\bullet T^*$, both invariant under the action of ${\rm SO}(T\oplus T^*)$. 

We take the first inner product from \cite{Gualtieri:2003dx}, see also references therein. Let $\sigma_1$ be the main antiautomorphism of the Clifford algebra, i.e. the one determined by the map $v_1 \otimes \ldots \otimes v_k \to v_k \otimes \ldots \otimes v_1$. It is not hard to check that it acts on elements of the Clifford algebra changing signs according to the degree of the corresponding element
\be\label{sigma}
\sigma_1(\omega) = \epsilon_1(p) \omega, \qquad \omega \in {\rm Cliff}(T\oplus T^*),
\ee
where $\epsilon_1(p) = 1$ when $p=0,1$ mod $4$ and $\epsilon_1(p) = -1$ when $p=2,3$ mod $4$. Then for $\Psi_1, \Psi_2\in \Lambda^\bullet T^*$ we define the inner product to be
\be\label{inner}
(\Psi_1, \Psi_2)_1 := \sigma_1(\Psi_1) \Psi_2 \Big|,
\ee
where the notation $\Big|$ means restriction to the top form in  $\Lambda^\bullet T^*$. The invariance of this inner product follows from 
\be
(c(v) \Psi_1, c(v) \Psi_2) = \sigma_1(c(v) \Psi_1) c(v) \Psi_2 \Big| = \sigma_1(\Psi_1) \sigma_1(c(v)) c(v) \Psi_2 \Big| \\ \nonumber
= G(v,v) \sigma_1(\Psi_1) \Psi_2 \Big|  = G(v,v) (\Psi_1, \Psi_2).
\ee
Taking into account that ${\rm Spin}(T\oplus T^*)$ sits inside the Clifford algebra as (\ref{spin-group}) we see that the inner product (\ref{inner}) is invariant under the identity component of Spin. The described inner product is symmetric when $n=0,1$ mod $4$, and anti-symmetric when $n=2,3$ mod $4$. 

We now describe another invariant inner product on spinors. We take this construction from \cite{Witt:2005sk}. It is given by the same construction (\ref{inner}), but with the involution $\sigma_1$ replaced with a different one $\sigma_2$. The involution $\sigma_2$ can again be described as changing signs according to degrees (\ref{sigma}), but now with $\epsilon_2(p)=1$ when $p=0,3$ mod $4$ and $\epsilon_2(p)=-1$ when $p=2,3$. So, the involutions $\sigma_1, \sigma_2$ differ by sign in what they do to odd elements of Clifford algebra. The second inner product is symmetric if $n=0,3$ mod $4$ and anti-symmetric if $n=1,2$ mod $4$. This is opposite symmetry property to (\ref{inner}) for odd $n$. 

It should also be noted that the spaces of even and odd forms in $\Lambda^\bullet T^*$ are null with respect to both inner products when $n$ is odd, and orthogonal to each other when $n$ is even. 

\subsection{The Weyl and Dirac Lagrangians}

We can now describe natural ${\rm SO}(n,n)$ invariant Lagrangians that can be constructed with the Dirac operator $D$ and the above invariant inner products on $\Lambda^\bullet T^*$. The construction is to take $(\Psi, D\Psi)_{1,2}$ with respect to one of the two described inner products. 

The construction is different depending on the parity of $n$. Let us describe the $n$ odd case first. In this case we can restrict the Lagrangian $(\Psi, D\Psi)_{1,2}$ to the space of even or odd forms in $\Lambda^\bullet T^*$, i.e. to the space of Weyl spinors. Indeed, if we decompose 
\be
\Psi = \Psi_+ + \Psi_-,
\ee
where $\Psi_+$ stands for even forms and $\Psi_-$ for odd, then we have
\be
(\Psi, D\Psi)_{1,2} = (\Psi_+, D\Psi_+)_{1,2} + (\Psi_-, D\Psi_-)_{1,2},
\ee
and so even and odd forms are not mixed by the kinetic term. They would be mixed if we wished to add to the Lagrangian terms like $(\Psi,\Psi)_{1,2}$, which are possible depending on which inner product is used in dimensions $n=1,3$ mod $4$. However, we will not be considering these Dirac mass terms, and restrict our attention to Weyl spinors, which we take to be given by even forms. Then no explicit mass terms can be written. It can then be checked that $(\Psi_+, D\Psi_+)_{1,2}$ vanishes modulo surface terms arising by integration by parts when $n=1$ mod $4$. This happens for both inner products. The Weyl Lagrangian $(\Psi_+, D\Psi_+)_{1,2}$ is only non-trivial for $n=3$ mod $4$, and in this case both inner products give the same result, modulo an overall sign. A proof of these statements is by explicit verification. 

Let us now consider the situation when $n$ is even. In this case the kinetic term mixes the even and odd forms
\be
(\Psi, D\Psi)_{1,2} = (\Psi_+, D\Psi_-)_{1,2} + (\Psi_-, D\Psi_+)_{1,2}.
\ee
It now does matter which inner product is chosen. It can be checked that the first described inner product gives a non-trivial Lagrangian for $n=0$ mod $4$ (and vanishes modulo surface terms for $n=2$ mod $4$), and the second product gives a non-trivial Dirac Lagrangian for $n=2$ mod $4$ (and vanishes for $n=0$ mod $4$). The Dirac mass term $(\Psi,\Psi)$ is only non-trivial for $n=0$ mod $4$.

To summarise, the Weyl Lagrangian $(\Psi_+, D\Psi_+)$ only exists for $n=3$ mod $4$, in which case it does not matter which inner product is used. The Dirac Lagrangian $(\Psi, D\Psi)$ exists for $n=0,2$ mod $4$, depending on which inner product is used. We will only consider the Weyl Lagrangian in this paper, and omit the subscript "plus" from $\Psi_+$ from now on.

\subsection{The second metric}

Let us now assume that on top of the geometric structure $F,G$ already present on $\cM$, we are given another endomorphism $J: J^2=\id$ and that is metric-compatible in the following sense
\be
G(JX,JY)=G(X,Y),
\ee
i.e. with no minus sign as in the case of $F$. Let us denote the eigenspaces of $J$ of eigenvalues $\pm 1$ as $U,\tilde{U}$, so that $T\cM = U\oplus\tilde{U}$. Then it is easy to show that the metric $G$ restricts to a non-degenerate metrics on $U,\tilde{U}$, and $U,\tilde{U}$ are G-orthogonal. Indeed, to show the orthogonality we take $X'\in U, Y''\in \tilde{U}$ and compute
\be
G(X', Y'') = G(J X', JY'') = - G(X', Y''),
\ee
and so $G(X', Y'')=0$. A similar computation shows that the restriction of $G$ to $U, \tilde{U}$ is non-degenerate. 

We can now parametrise such endomorphisms $J$ by what in the double field theory context is usually referred to as the generalised metric. Let us see how this can be done. First, each of the spaces $U,\tilde{U}$, being of same dimension as $T$, can be described as a graph of some map $T\to T^*$, where we again identified $\tilde{T}=T^*$. Each such map is a rank two tensor, and let us denote by $g$ its symmetric part, and by $b$ its anti-symmetric part. So, the space $U$ can be parametrised as consisting of elements
\be\label{U}
\xi + (b+g)\xi \in U,
\ee
where $b,g\in T^*\otimes T^*$ are some anti-symmetric and symmetric tensors. A moment of reflection shows that the space that is $G$-orthogonal to $U$ is then
\be\label{Ut}
\xi + (b-g)\xi \in \tilde{U},
\ee
where the same tensors are used. The restriction of the metric (\ref{G}) to $U$ is then
\be
G(\xi_1+ (b+g)\xi_1,\xi_2 + (b+g)\xi_2) = g(\xi_1,\xi_2),
\ee
and the restriction to $\tilde{U}$ is minus this. 

Let us now use the data $g,b$ to construct an endomorphism of $T\oplus T^*$ that squares to identity, is metric compatible and whose eigenspaces are as described above. This endomoprhism is explicitly given by
\be\label{J-plus}
J= \left( \begin{array}{cc} - g^{-1} b & g^{-1}  \\ g- b g^{-1} b & b g^{-1} \end{array} \right).
\ee
It is easy to see that this endomorphism is designed to square to the identity
\be
J^2=\left( \begin{array}{cc} \id  & 0  \\ 0 & \id \end{array} \right).
\ee
The endomorphism $J$ is also metric-compatible $G(J\cdot,J\cdot)=G(\cdot,\cdot)$. To check this, we must compute 
\be
J X= \left( \begin{array}{cc} - g^{-1} b & g^{-1}  \\ g- b g^{-1} b & b g^{-1} \end{array} \right)\left( \begin{array}{c} \xi \\ \eta \end{array} \right) = \left( \begin{array}{c} - g^{-1} b\xi + g^{-1} \eta \\ (g-bg^{-1} b)\xi + b g^{-1} \eta \end{array} \right).
\ee
Let us now pair $JX_1,JX_2$. We have
\be
G(JX_1,JX_2) = \frac{1}{2} ( \xi_1 (g- bg^{-1} b) - \eta_1 g^{-1} b) ( - g^{-1} b \xi_2 + g^{-1} \eta_2) + \\
 \frac{1}{2}( \xi_2 (g- bg^{-1} b) - \eta_2 g^{-1} b) ( - g^{-1} b \xi_1 + g^{-1} \eta_1),
\ee
where some transposes where taken and minus signs from $b^T=-b$ introduced. Opening up the brackets and seeing the cancellations one verifies $G(JX_1,JX_2) = G(X,Y)$. Let us also see that the eigenspaces of $J$ are as described above. The eigenvector equation for eigenvalue $+1$ is
\be
JX=X \quad \Rightarrow \quad - g^{-1} b \xi + g^{-1} \eta = \xi \Rightarrow \quad \eta=(b+g)\xi.
\ee
Then the second of the arising equations $(g- b g^{-1} b)\xi + b g^{-1} \eta = \eta$ is automatically satisfied. Thus, we learn that the eigenvectors of eigenvalue $+1$ are of the form (\ref{U}) and those of eigenvalue $-1$ are of the form (\ref{Ut}). So, $J$ given by (\ref{J-plus}) is indeed the required endomorphism. 

\subsection{Another basis for $V$}

We thus assume that in addition to data $F,G$ there is some mechanism that gives rise to an endomorphism $J$ as described above, and thus to tensors $g,b\in T^*\otimes T^*$. A possible origin  of such mechanism will be discussed in the last section. For simplicity we assume that $b=0$ in what follows. We will also usually assume that $g$ is the flat Riemannian signature metric in $T$. 

With this assumption a vector $\xi+\eta\in V$ can be decomposed into its $U,\tilde{U}$ parts as
\be
\xi + \eta = u + g u + \tilde{u} - g \tilde{u}, \qquad u= \frac{1}{2}( \xi + g^{-1} \eta), \qquad \tilde{u}= \frac{1}{2}( \xi - g^{-1} \eta).
\ee
The inverse of this transformation, in matrix form
\be\label{u-coord}
\left(\begin{array}{c} \xi \\ \eta \end{array} \right) = \left(\begin{array}{cc} \id & \id \\ g & -g \end{array} \right) \left(\begin{array}{c} u \\ \tilde{u} \end{array} \right).
\ee
In this basis the metric is
\be
G((u_1,\tilde{u}_1),(u_2,\tilde{u}_2)) = g(u_1,u_2)- g(\tilde{u}_1,\tilde{u}_2),
\ee
and the Lie algebra is represented by matrices
\be\label{M-UV-basis}
\frac{1}{2} \left(\begin{array}{cc} \id & g^{-1} \\ \id & -g^{-1} \end{array} \right) \left( \begin{array}{cc} A & \beta \\ B & -A^T \end{array}\right) \left(\begin{array}{cc} \id & \id \\ g & -g \end{array} \right) \\ \nonumber
=\frac{1}{2} \left( \begin{array}{cc} A-g^{-1} A^T g + g^{-1} B+ \beta g & A+g^{-1} A^Tg +g^{-1} B-\beta g\\ A+g^{-1} A^T g -(g^{-1} B-\beta g) & A-g^{-1} A^Tg -(g^{-1} B+\beta g) \end{array}\right),
\ee
which have $g$-anti-symmetric matrices on the diagonal, and have the off-diagonal blocks $g$-transpose of each other. When $g_{ij}=\delta_{ij}$ the factors of $g, g^{-1}$ can be simply removed from this formula, with positions of indices adjusted appropriately. We will need this result below when we describe embedding of various subgroups into ${\rm SO}(n,n)$.

\section{The case of ${\rm SO}(2,2)$}
\label{sec:22}

In this section we work out the spinor representations of ${\rm SO}(2,2)$ and explicitly verify that the Dirac operator as described in (\ref{dirac}) is the usual Dirac operator for the split signature metric in four dimensions. Readers that do not need such an explicit verification can skip this section.

\subsection{Spinor representations of ${\rm SO}(2,2)$}

Let us see explicitly how the spinor representations of ${\rm SO}(2,2)$ are differential forms in $\R^2$. To this end, we introduce a pair of creating annihilation operators $a_1, (a^1)^\dagger$ and $a_2, (a^2)^\dagger$, with the usual anti-commutation relations $a_i (a^j)^\dagger + (a^j)^\dagger a_i = \delta_i^j$ and all other pairs anti-commuting. We can then consider the following operators
\be
H =  a_1 (a^1)^\dagger - a_2 (a^2)^\dagger, \qquad E_+ = a_1 (a^2)^\dagger, \qquad E_- = a_2 (a^1)^\dagger.
\ee
It is easy to check that the following ${\mathfrak sl}(2)$ commutation relations hold
\be\label{sl2-alg}
[E_+, E_-]=H, \qquad [H,E_\pm]=\pm 2 E_\pm.
\ee

This gives us one copy of ${\mathfrak sl}(2)$ Lie algebra. One can form the second copy of ${\mathfrak sl}(2)$ in the following way
\be
\bar{H} = a_1 (a^1)^\dagger + a_2 (a^2)^\dagger - 1\equiv a_1 (a^1)^\dagger - (a^2)^\dagger a_2, \qquad \bar{E}_+ = a_1 a_2, \qquad \bar{E}_- = (a^2)^\dagger (a^1)^\dagger.
\ee
Again we get the usual ${\mathfrak sl}(2)$ commutation relations
\be
[\bar{E}_+,\bar{E}_-] = \bar{H}, \qquad [\bar{H},\bar{E}_\pm]=\pm 2 \bar{E}_\pm.
\ee
And it is not hard to check that all barred operators commute with unbarred ones, so we have two commuting copies of ${\mathfrak sl}(2)$. If we do this construction over reals we get an explicit realisation of the Lie algebra of ${\mathfrak so}(2,2)$ as two commuting Lie algebras ${\mathfrak sl}(2,\R)$.

Let us now discuss its action on spinors. The Weyl representations are formed by forms of even and odd degrees. The forms of odd degree are spanned by $dx^1, dx^2$. The action of the first copy of ${\mathfrak sl}(2)$ is as follows
\be\label{22-action}
H dx^2 = (a_1 (a^1)^\dagger - a_2 (a^2)^\dagger)dx^2 = dx^2, \qquad H dx^1 = (a_1 (a^1)^\dagger - a_2 (a^2)^\dagger) dx^1 = - dx^1, \\ \nonumber
E_- dx^2 = a_2 (a^1)^\dagger dx^2 = - dx^1, \qquad E_+  dx^1 = a_1 (a^2)^\dagger dx^1 = - dx^2,
\ee
while the second copy acts trivially on these states. 

The algebra (\ref{sl2-alg}) is realised the by the matrices 
\be
E_+ = \left( \begin{array}{cc} 0 & 1 \\ 0 & 0 \end{array} \right), \quad E_- = \left( \begin{array}{cc} 0 & 0 \\ 1 & 0 \end{array} \right), \quad H = \left( \begin{array}{cc} 1 & 0 \\ 0 & -1 \end{array} \right).
\ee
A comparison of this with the action (\ref{22-action}) then fixes the identification of forms with 2-component column spinors up to an overall sign, which we choose as follows
\be\label{22-ident-odd}
\left( \begin{array}{c} \bar{\alpha} \\ \bar{\beta} \end{array} \right) = -\bar{\alpha} dx^2 + \bar{\beta} dx^1.
\ee

The even degree forms are spanned by $1$ and $dx^1 dx^2$. The first copy of ${\mathfrak sl}(2)$ acts trivially, while the action of the second copy is
\be
\bar{H}  1= ( a_1 (a^1)^\dagger - (a^2)^\dagger a_2) 1 = 1, \qquad \bar{H}  dx^1 dx^2 = ( a_1 (a^1)^\dagger - (a^2)^\dagger a_2) dx^1 dx^2 = - dx^1 dx^2, \\ \nonumber
\bar{E}_- 1 = (a^2)^\dagger (a^1)^\dagger  1 = - dx^1 dx^2, \qquad \bar{E}_+ dx^1 dx^2 = a_1 a_2 dx^1 dx^2 = - 1.
\ee
The identification with 2-column spinors that we choose for this copy is 
\be\label{22-ident-even}
 \left( \begin{array}{c} \alpha \\ \beta \end{array} \right) =-\alpha + \beta dx^1 dx^2.
 \ee

\subsection{The Dirac operator on $\R^{2,2}$}

We start with $\R^{(2,2)}$ with metric in the diagonal form
\be
ds^2 = (du^1)^2 + (du^2)^2 - (d\tilde{u}^1)^2 - (d\tilde{u}^2)^2.
\ee
In terms of coordinates
\be
x^{1,2}= u^{1,2}+\tilde{u}^{1,2}, \qquad \tilde{x}_{1,2}= u^{1,2}-\tilde{u}^{1,2}
\ee
the metric is
\be
ds^2 = dx^1 d\tilde{x}_1 + dx^2 d\tilde{x}_2.
\ee

Consider the real $2\times 2$ matrix
\be
{\bf u}^{AA'} = \left( \begin{array}{cc} u^1 + \tilde{u}^1 & u^2 + \tilde{u}^2 \\ u^2 - \tilde{u}^2 & -u^1 + \tilde{u}^1 \end{array} \right) = \left( \begin{array}{cc} x^1 & x^2  \\ \tilde{x}_2 & - \tilde{x}_1 \end{array} \right).
\ee
Here $A,A'=1,2$ are 2-component spinor indices. For the matrix ${\bf u}^{AA'}$ the index $A'$ enumerates columns and $A$ enumerates rows. The determinant of the above matrix is minus the squared interval. 

Let us now construct the related matrices ${\bf u}_A{}^{A'}$ and ${\bf u}_{A'}{}^A$. The first one is given by $-\epsilon {\bf u}$, and the second one by $-\epsilon {\bf u}^T$, where 
\be\label{epsilon}
\epsilon=  \left( \begin{array}{cc} 0 & 1 \\ -1 & 0 \end{array} \right)
\ee
is the metric in the space of spinors, with the conventions being
\be
\lambda^A \mu_A = - \lambda_A \epsilon^{AB} \mu_B = - \lambda^T \epsilon \mu,
\ee
so that $(\epsilon \lambda)^T$ is the row representing $\lambda^A$. Then $\lambda^A \mu_A$ is the usual matrix product of a row and a column. We have
\be
{\bf u}_{A}{}^{A'} = \left( \begin{array}{cc} u^2 - \tilde{u}^2 & -u^1 + \tilde{u}^1  \\ - u^1 - \tilde{u}^1 &  u^2 + \tilde{u}^2 \end{array} \right)= \left( \begin{array}{cc} \tilde{x}_2 & -\tilde{x}_1 \\ -x^1  & - x^2 \end{array} \right),\\ \nonumber
{\bf u}_{A'}{}^A = \left( \begin{array}{cc} u^2 + \tilde{u}^2 & - u^1 + \tilde{u}^1   \\ -u^1 - \tilde{u}^1 & -u^2 + \tilde{u}^2   \end{array} \right)= \left( \begin{array}{cc} x^2  & -\tilde{x}_1   \\ - x^1& -\tilde{x}_2  \end{array} \right).
\ee
We have
\be
{\bf u}_{A'}{}^A {\bf u}_{A}{}^{B'} = |x|^2 \id_{A'}{}^{B'}, \qquad {\bf u}_A{}^{A'} {\bf u}_{A'}{}^B = |x|^2 \id_A{}^B, \qquad
 |x|^2 = x^1 \x_1 + x^2 \x_2.
\ee
We can then form the $4\times 4$ matrix
\be
\backslash\!\!\! u = \left( \begin{array}{cc} 0 & {\bf u}_A{}^{A'} \\ {\bf u}_{A'}{}^A & 0 \end{array} \right)
\ee
that acts on 4-component Dirac spinors and satisfies the desired Clifford algebra relation
\be
\backslash\!\!\! u \backslash\!\!\! u = |x|^2 \id.
\ee
This construction gives us the two chiral Dirac operators 
\be
\partial_{A}{}^{A'} \equiv \partial^T = \left( \begin{array}{cc} \partial/\partial u^2  - \partial/\partial \tilde{u}^2 & - \partial/\partial u^1  + \partial/\partial \tilde{u}^1 \\  -\partial/\partial u^1  - \partial/\partial \tilde{u}^1 & -\partial/\partial u^2  - \partial/\partial \tilde{u}^2  \end{array} \right) = 2\left( \begin{array}{cc} \partial/\partial \tilde{x}_2 & - \partial/\partial \tilde{x}_1 \\ -\partial/\partial x^1 & -\partial/\partial x^2 \end{array}\right), \\ \nonumber
\partial_{A'}{}^A \equiv \partial = \left( \begin{array}{cc} \partial/\partial u^2  + \partial/\partial \tilde{u}^2 & -\partial/\partial u^1  + \partial/\partial \tilde{u}^1  \\ - \partial/\partial u^1  - \partial/\partial \tilde{u}^1 & -\partial/\partial u^2  + \partial/\partial \tilde{u}^2   \end{array} \right)= 2\left( \begin{array}{cc} \partial/\partial x^2 & -\partial/\partial \tilde{x}_1   \\ - \partial/\partial x_1 & -\partial/\partial \tilde{x}_2   \end{array}\right).
\ee
We can write these operators more compactly as
\be
\partial^T  = 2\left( \begin{array}{cc} \tilde{\partial}^2  & - \tilde{\partial}^1  \\ -\partial_1 & -\partial_2 \end{array} \right), \qquad \partial = 2\left( \begin{array}{cc} \partial_2 & -\tilde{\partial}^1   \\  -\partial_1  & -\tilde{\partial}^2  \end{array} \right).
\ee
These are the two chiral Dirac operators 
\be
\partial^T : S_- \to S_+, \qquad \partial: S_+ \to S_-.
\ee
The action on a primed spinor in $S_-$ is
\be
\partial^T \left( \begin{array}{c} \bar{\alpha} \\ \bar{\beta} \end{array} \right) =2 \left( \begin{array}{c} \tilde{\partial}^2 \bar{\alpha} - \tilde{\partial}^1\bar{\beta} \\ -\partial_1 \bar{\alpha} -\partial_2 \bar{\beta}  \end{array} \right),
\ee
and on unprimed one we have
\be
\partial \left( \begin{array}{c} \alpha \\ \beta \end{array} \right) = 2\left( \begin{array}{c} \partial_2 \alpha - \tilde{\partial}^1 \beta \\ - \partial_1 \alpha - \tilde{\partial}^2 \beta  \end{array} \right).
\ee

\subsection{The Dirac operator as the exterior derivative}

We now show verify that the Dirac operator on $\R^{2,2}$ is essentially the exterior derivative operator appropriately interpreted. Thus, we identify the space of primed spinors with the space of odd forms as in (\ref{22-ident-odd}). The exterior derivative operator is then
\be
D( -\bar{\alpha} dx^2 + \bar{\beta} dx^1) = -\partial_1 \bar{\alpha} dx^1 dx^2 + \partial_2 \bar{\beta} dx^2 dx^1 - \tilde{\partial}^2 \bar{\alpha} d\x_2 dx^2 + \tilde{\partial}^1 \bar{\beta}  d\x_1 dx^1 = \\ \nonumber
( -\partial_1 \bar{\alpha}-\partial_2 \bar{\beta} ) dx^1 dx^2 -  (\tilde{\partial}^2 \bar{\alpha}-\tilde{\partial}^1 \bar{\beta}) .
\ee
Here the differentials $d\x_1, d\x_2$ are interpreted as annihilation operators that can act on $dx^1, dx^2$ non-trivially, and only the terms giving non-zero contribution have been kept. We used a different letter for the exterior derivative to signify the fact that $D^2\not=0$. 

We can write the above result as
\be
D\left( \left( \begin{array}{cc} - dx^2 & dx^1 \end{array}\right) \left( \begin{array}{c} \bar{\alpha} \\ \bar{\beta} \end{array}\right) \right) =  \left( \begin{array}{cc} - 1 & dx^1 dx^2 \end{array}\right) \frac{1}{2} \partial^T \left( \begin{array}{c} \bar{\alpha} \\ \bar{\beta} \end{array} \right) .
\ee
Taking into account (\ref{22-ident-odd}), (\ref{22-ident-even}) we see that $D$ indeed gives the correct chiral Dirac operator when it acts on odd forms. 

We can similarly compute the action of $D$ on even forms
\be
D(-\alpha + \beta dx^1 dx^2) = - \partial_1 \alpha dx^1 - \partial_2 \alpha dx^2 + \tilde{\partial}^1 \beta d\x_1 dx^1 dx^2 + \tilde{\partial}^2 \beta d\x_2 dx^1 dx^2  \\ \nonumber
= - (\partial_2 \alpha - \tilde{\partial}^1 \beta ) dx^2+ (-\partial_1 \alpha- \tilde{\partial}^2 \beta)  dx^1 .
\ee 
We see that
\be
D\left(\left( \begin{array}{cc} - 1 & dx^1 dx^2 \end{array}\right)\left( \begin{array}{c} \alpha \\ \beta \end{array} \right) \right) = \left( \begin{array}{cc} - dx^2 & dx^1 \end{array}\right) \frac{1}{2} \partial \left( \begin{array}{c} \alpha \\ \beta \end{array} \right).
\ee
This verifies that the exterior derivative operator $D$, interpreted in the sense of Clifford multiplication, matches the Dirac operator on $\R^{2,2}$ In particular, this shows that $D^2=(1/4)\Delta$, where $\Delta$ is the Laplacian on $\R^{(2,2)}$. 

\section{Case of ${\rm SO}(3,3)$}
\label{sec:33}

The group ${\rm SO}(3,3)$ is the smallest of ${\rm SO}(n,n)$ groups that contains the Lorentz group ${\rm SO}(3,1)$. 

\subsection{Embedding of ${\rm SO}(3,1)\times{\rm SO}(2)$}

To select a copy of the Lorentz group sitting inside ${\rm SO}(3,3)$ we pass to the $u,\tilde{u}$ coordinates (\ref{u-coord}) that make the metric diagonal. We do this by choosing the metric $g$ to be a flat metric of signature all plus. Using this metric we can lower-raise the indices of $x^i, \tilde{x}_i$, and also of matrices $\beta^{ij}, B_{ij}, A_i{}^j$. We will write all coordinates with indices down so that $a_i^\dagger = dx_i$ and $a_i= d\x_i$, and matrices with indices up. 

We want to embed the Lorentz group ${\rm SO}(3,1)$ in the "diagonal" way into ${\rm SO}(3,3)$. The Lie algebra of ${\rm SO}(n,n)$ in the basis in which the metric is diagonal is formed by matrices of the form (\ref{M-UV-basis}). We assume $g=\delta$ everywhere. To describe a copy of ${\rm SO}(3,1)$ inside let us start with the rotations subgroup. This is embedded into the upper-left corner of the $6\times 6$ matrix (\ref{M-UV-basis}). So we want $A+A^T+ B-\beta=0, A-A^T - (B+\beta)=0$ and so $A=\beta, B=\beta$, which in the familiar $\xi,\eta$ basis corresponds to matrices of the form
\be\label{rotations}
\left( \begin{array}{cc} \beta & \beta \\ \beta & \beta \end{array}\right),
\ee
with $\beta$ being the usual $3\times 3$ matrices representing rotations of the first three coordinates. This leads to the following matrices
\be
 K_1=-\frac{1}{2} \left(
\begin{array}{cccccc}
 0 & 0 & 0  & 0 & 0 & 0  \\
 0 & 0 & 1  & 0 & 0 & 1  \\
 0 & -1 & 0  & 0 & -1 & 0  \\
 0 & 0 & 0  & 0 & 0 & 0  \\
 0 & 0 & 1  & 0 & 0 & 1  \\
 0 & -1 & 0  & 0 & -1 & 0  \\
\end{array}
\right),
K_2=-\frac{1}{2} \left(
\begin{array}{cccccccc}
 0 & 0 & -1  & 0 & 0 & -1  \\
 0 & 0 & 0  & 0 & 0 & 0  \\
 1 & 0 & 0  & 1 & 0 & 0  \\
 0 & 0 & -1  & 0 & 0 & -1  \\
 0 & 0 & 0  & 0 & 0 & 0  \\
 1 & 0 & 0  & 1 & 0 & 0  \\
\end{array}
\right), \\ \nonumber
K_3=-\frac{1}{2}\left(
\begin{array}{cccccccc}
 0 & 1 & 0  & 0 & 1 & 0  \\
 -1 & 0 & 0  & -1 & 0 & 0  \\
 0 & 0 & 0  & 0 & 0 & 0  \\
 0 & 1 & 0  & 0 & 1 & 0  \\
 -1 & 0 & 0  & -1 & 0 & 0  \\
 0 & 0 & 0  & 0 & 0 & 0  \\
\end{array}
\right),
\ee
where we included the prefactors in order to get the correct normalisation, see below. 

Let us now discuss the boosts. We choose this to mix the coordinate $\tilde{u}^3$ with the coordinates $u^i$. They are thus represented in the $u,\tilde{u}$ basis by matrices (\ref{M-UV-basis}) with zero on the diagonal $A-A^T\pm(B+\beta)=0$ , and with the off-diagonal block equal to $(A+A^T+B-\beta)_{ij} = \delta_{is} \delta_{j3}$, where $s=1,2,3$. These leads to $A$ being a symmetric matrix, and $B=-\beta$, and the following matrices in the $\eta,\xi$ representation:
\be
P_1= \frac{1}{2} \left(
\begin{array}{cccccc}
  0 & 0 & 1  & 0 & 0 & -1  \\
 0 & 0 & 0  & 0 & 0 & 0  \\
 1 & 0 & 0  & 1 & 0 & 0  \\
 0 & 0 & 1  & 0 & 0 & -1  \\
 0 & 0 & 0  & 0 & 0 & 0  \\
 -1 & 0 & 0  & -1 & 0 & 0  \\
\end{array}
\right),
P_2=\frac{1}{2}\left(
\begin{array}{cccccc}
  0 & 0 & 0  & 0 & 0 & 0  \\
 0 & 0 & 1  & 0 & 0 & -1  \\
 0 & 1 & 0  & 0 & 1 & 0  \\
 0 & 0 & 0  & 0 & 0 & 0  \\
 0 & 0 & 1  & 0 & 0 & -1  \\
 0 & -1 & 0  & 0 & -1 & 0  \\
\end{array}
\right), \\ \nonumber
P_3=\left(
\begin{array}{cccccc}
 0 & 0 & 0  & 0 & 0 & 0  \\
 0 & 0 & 0  & 0 & 0 & 0  \\
 0 & 0 & 1  & 0 & 0 & 0  \\
 0 & 0 & 0  & 0 & 0 & 0  \\
 0 & 0 & 0  & 0 & 0 & 0  \\
 0 & 0 & 0  & 0 & 0 & -1  \\
\end{array}
\right).
\ee

The above set of matrices $K_i, P_i$ can then be checked to have the usual Lorentz group commutation relations 
\be\label{Lorentz}
[K_i,K_j]=\epsilon_{ijk} K_k, \qquad [K_i,P_j]=\epsilon_{ijk}P_k, \qquad [P_i,P_j]=-\epsilon_{ijk}K_k.
\ee
These matrices act on differential forms as (\ref{action-M-forms}), and so correspond to the following set of operators on differential forms
\be\label{Lorentz-K}
K_1 = -\frac{1}{2}\left( a_2 a_3^\dagger - a_3 a_2^\dagger + a_2 a_3 + a_2^\dagger a_3^\dagger\right),
K_2 = -\frac{1}{2}\left( a_3 a_1^\dagger - a_1 a_3^\dagger + a_3 a_1 + a_3^\dagger a_1^\dagger\right),\\ \nonumber
K_3 = -\frac{1}{2}\left( a_1 a_2^\dagger - a_2 a_1^\dagger + a_1 a_2 + a_1^\dagger a_2^\dagger\right), 
\ee
which are all anti-Hermitian, and
\be\label{Lorentz-P}
P_1= \frac{1}{2}\left( a_1 a_3^\dagger + a_3 a_1^\dagger - a_1 a_3 + a_1^\dagger a_3^\dagger\right),
P_2= \frac{1}{2}\left( a_2 a_3^\dagger + a_3 a_2^\dagger - a_2 a_3 + a_2^\dagger a_3^\dagger\right),\\ \nonumber
P_3= \frac{1}{2}\left( a_3 a_3^\dagger - a_3^\dagger a_3 \right),
\ee
which are all Hermitian. 

We now work out the similar embedding of ${\rm SO}(2)$ subgroup that mixes the $\tilde{u}^1, \tilde{u}^2$ coordinates. In the $u,\tilde{u}$ basis this corresponds to matrices with off-diagonal blocks equal to zero, and thus $A+A^T+(B-\beta)=0$, and with the upper-diagonal block equal to zero $A-A^T+B+\beta=0$. This gives $B=-A, \beta=A^T$. Thus, these are matrices of the form
\be\label{rotations-dual}
\left( \begin{array}{cc} -\beta & \beta \\ \beta & -\beta \end{array}\right).
\ee
The particular rotation that we are after is represented by the following matrix
\be
C=\frac{1}{2}\left(
\begin{array}{cccccc}
 0 & 1 & 0  & 0 & -1 & 0  \\
 -1 & 0 & 0  & 1 & 0 & 0  \\
 0 & 0 & 0  & 0 & 0 & 0  \\
 0 & -1 & 0  & 0 & 1 & 0  \\
 1 & 0 & 0  & -1 & 0 & 0  \\
 0 & 0 & 0  & 0 & 0 & 0  \\
\end{array}
\right),
\ee
which corresponds to the operator
\be
C= \frac{1}{2} \left( a_1 a_2^\dagger - a_2 a_1^\dagger - a_1 a_2 - a_1^\dagger a_2^\dagger\right).
\ee
This operator is anti-Hermitian, as is appropriate for a rotation. It can be checked that the matrix $C$ commutes with $K_i, P_i$ as it should. 

\subsection{Change of basis}

To describe the action of all the operators on differential forms, we introduce the complex linear combinations 
\be\label{dm-dbarm}
dm=\frac{1}{\sqrt{2}}( dx_1 - \im dx_2), \qquad d\bar{m}=\frac{1}{\sqrt{2}}( dx_1 + \im dx_2).
\ee
We then define a new set of creation and annihilation operators, corresponding to creation-annihilation of $m,\bar{m}$
\be\label{am-mbar}
a_m := \frac{1}{\sqrt{2}}( a_1 -\im a_2), \qquad a_{\bar{m}} := \frac{1}{\sqrt{2}}( a_1 +\im a_2), \\ \nonumber
a_m^\dagger := \frac{1}{\sqrt{2}}( a_1^\dagger -\im a_2^\dagger), \qquad a_{\bar{m}}^\dagger := \frac{1}{\sqrt{2}}( a_1^\dagger +\im a_2^\dagger).
\ee
The anti-commutation relations are now
\be
a_m a_{\bar{m}}^\dagger + a_{\bar{m}}^\dagger a_m = 1, \qquad a_{\bar{m}} a_m^\dagger + a_m^\dagger a_{\bar{m}} = 1, 
\ee
while 
\be
a_m a_m^\dagger + a_m^\dagger a_m = 0, \qquad a_{\bar{m}} a_{\bar{m}}^\dagger + a_{\bar{m}}^\dagger a_{\bar{m}}=0.
\ee
This is of course just the usual Clifford algebra relations corresponding to the fact that the metric in $1,2$ plane in the new basis is $dm\otimes d\bar{m} + d\bar{m}\otimes dm$. In the new basis, the operator $K_3$ takes the following form
\be
K_3 = -\frac{\im}{2} \left( a_{\bar{m}} a_m^\dagger - a_m a_{\bar{m}}^\dagger + a_{\bar{m}} a_m + a_{\bar{m}}^\dagger a_m^\dagger \right).
\ee
This immediately gives the eingestates of $K_3$
\be\label{K3-action}
K_3 dm = \frac{\im}{2} dm, \quad K_3 d\bar{m} = -\frac{\im}{2}d\bar{m} , \quad 
K_3 (1\pm dm d\bar{m}) = \pm \frac{\im}{2} (1\pm dm d\bar{m}).
\ee 

We also need the operator $C$ in the new basis
\be
C= \frac{\im}{2} \left( a_{\bar{m}} a_m^\dagger - a_m a_{\bar{m}}^\dagger - a_{\bar{m}} a_m - a_{\bar{m}}^\dagger a_m^\dagger \right),
\ee
with eigenstates being
\be\label{C-action}
C dm = -\frac{\im}{2} dm, \quad C d\bar{m} = \frac{\im}{2}d\bar{m} , \quad 
C (1\pm dm d\bar{m}) = \pm \frac{\im}{2} (1\pm dm d\bar{m}).
\ee 

We also list the eigenstate of $P_3$
\be
P_3 dx_3 = - \frac{1}{2} dx_3, \qquad P_3 1 = \frac{1}{2} 1.
\ee

It is also convenient to introduce the complex linear combinations $K_1\pm \im K_2$ and $P_1\pm \im P_2$. We can rewrite these operators in the new basis as
\be\label{K-pm}
\frac{1}{\sqrt{2}}(K_1 -\im K_2) = -\frac{\im}{2} ( a_m a_3^\dagger - a_3 a_m^\dagger + a_m a_3 - a_3^\dagger a_m^\dagger), \\ \nonumber
\frac{1}{\sqrt{2}}(K_1 +\im K_2) = \frac{\im}{2} ( a_{\bar{m}} a_3^\dagger - a_3 a_{\bar{m}}^\dagger + a_{\bar{m}} a_3 - a_3^\dagger a_{\bar{m}}^\dagger),
\ee
and
\be\label{P-pm}
\frac{1}{\sqrt{2}}(P_1 -\im P_2) = \frac{1}{2} ( a_m a_3^\dagger + a_3 a_m^\dagger - a_m a_3 + a_m^\dagger a_3^\dagger ), \\ \nonumber
\frac{1}{\sqrt{2}}(P_1 +\im P_2) = \frac{1}{2} ( a_{\bar{m}} a_3^\dagger + a_3 a_{\bar{m}}^\dagger - a_{\bar{m}} a_3 + a_{\bar{m}}^\dagger a_3^\dagger ).
\ee
Finally, we introduce the usual self-dual/anti-self-dual combinations
\be\label{E-pm}
E_-:=\frac{1}{2}(K_1-\im K_2) +\frac{\im}{2}(P_1-\im P_2) = \frac{\im}{\sqrt{2}} a_3 (a_m+a_m^\dagger), \\ \nonumber
E_+:=\frac{1}{2}(K_1+\im K_2) +\frac{\im}{2} (P_1+\im P_2) = \frac{\im}{\sqrt{2}} (a_{\bar{m}}+a_{\bar{m}}^\dagger) a_3^\dagger,
\ee
and
\be\label{bar-E-pm}
\bar{E}_-:=\frac{1}{2}(K_1-\im K_2) -\frac{\im}{2} (P_1-\im P_2) = \frac{\im}{\sqrt{2}} a_3^\dagger (a_m+a_m^\dagger), \\ \nonumber
\bar{E}_+:=\frac{1}{2}(K_1+\im K_2) -\frac{\im}{2} (P_1+\im P_2) = \frac{\im}{\sqrt{2}} (a_{\bar{m}}+a_{\bar{m}}^\dagger) a_3.
\ee

\subsection{$2\times 2$ matrix realisation of the Lorentz Lie algebra}

For reference, we give here the $2\times 2$ matrix realisation of the Lie algebra ${\mathfrak so}(3,1)$. In this realisation the generators
\be
L^i = \frac{1}{2}(K^i -\im P^i), \qquad R^i = \frac{1}{2}(K^i +\im P^i)
\ee
are given by
\be
L^i = -\frac{\im}{2} \sigma^i, \qquad R^i = \frac{\im}{2} \sigma^i
\ee
respectively, where $\sigma^i$ are the usual Pauli matrices. And so we have
\be\label{E-2}
E_- = R^1 -\im R^2 = \im \left( \begin{array}{cc} 0 & 0 \\ 1 & 0 \end{array}\right), \qquad
E_+ = R^1 +\im R^2 = \im \left( \begin{array}{cc} 0 & 1 \\ 0 & 0 \end{array}\right),
\ee
and 
\be\label{bar-E-2}
\bar{E}_- = L^1 -\im L^2 = -\im \left( \begin{array}{cc} 0 & 0 \\ 1 & 0 \end{array}\right), \qquad
\bar{E}_+ = L^1 +\im L^2 = -\im \left( \begin{array}{cc} 0 & 1 \\ 0 & 0 \end{array}\right),
\ee
which shows that $E_-, \bar{E}_-$ are lowering and $E_+, \bar{E}_+$ are raising operators, as the notation suggests.

\subsection{Eigenstates}

We now consider the Weyl representations of ${\rm SO}(3,3)$, which are the spaces of even and odd forms in $\R^3$. We want to exhibit a basis in this space that diagonalises the operators $K_3\pm \im P_3$ and $C$. We have 8-dimensional space of forms on $\R^3$ that is spanned by forms $m,\bar{m}, (1\pm m\bar{m})$ and the same forms times $dx_3$. They are all eigenstates of $K_3, P_3$, and also of $C$, and so we just have to divide the states into two groups that transform non-trivially under $K_3 +\im P_3$ and trivially under $K_3 -\im P_3$ and vice versa. 

Let us first describe the even forms. The corresponding Weyl representation consists of four real states, which can be represented as two complex and their complex conjugates. The states that transform non-trivially under $K_3+\im P_3$ are
\be\label{primed}
 \frac{1}{\sqrt{2}}(1+dm d\bar{m}) \qquad {\rm eigenvalue\,\, of\,\, } K_3 +\im P_3 \quad +\im, \\ \nonumber
d\bar{m} dx_3  \qquad {\rm eigenvalue\,\, of \,\,} K_3 +\im P_3 \quad -\im,
\ee
and they thus form a single primed spinor of ${\rm SO}(3,1)$, and the states that transform non-trivially under $K_3-\im P_3$ are
\be\label{unprimed}
dm dx_3 \qquad {\rm eigenvalue\,\, of \,\,} K_3 -\im P_3 \quad +\im, \\ \nonumber
\frac{1}{\sqrt{2}} (1-dm d\bar{m})  \qquad {\rm eigenvalue\,\, of \,\,} K_3 -\im P_3 \quad -\im.
\ee
The factors of $1/\sqrt{2}$ are introduced for future convenience. These two ${\rm SO}(3,1)$ spinors are also eigenstates of $C$, with the primed spinor (\ref{primed}) being eigenstate of eigenvalue $+\im/2$, and the unprimed spinor (\ref{unprimed}) having the eigenvalue $-\im/2$. So, the content of the Weyl representation of ${\rm SO}(3,3)$ on even forms in $\R^3$ is a single electrically charged 2-component Weyl fermion plus its complex conjugate ${\bf 4}_\R={\bf 2}_\C$.

We can also check that the states (\ref{primed}), (\ref{unprimed}) transform correctly under the operators (\ref{E-pm}), (\ref{bar-E-pm}). We have for the primed spinor
\be\label{action-pr-even}
E_- d\bar{m} dx_3 = \im \frac{1}{\sqrt{2}} (1+dm d\bar{m}), \qquad E_+ \frac{1}{\sqrt{2}} (1+dm d\bar{m}) = \im d\bar{m} dx_3,
\ee
and for the unprimed one
\be
\bar{E}_- \frac{1}{\sqrt{2}} (1-dm d\bar{m}) = -\im dm dx_3, \qquad \bar{E}_+ dm dx_3 = -\im \frac{1}{\sqrt{2}} (1-dm d\bar{m}).
\ee
Comparison with (\ref{bar-E-2}) then shows that we should identify a linear combination of these forms with the following 2-component spinor
\be\label{ident-un-even}
\left( \begin{array}{cc}  \frac{1}{\sqrt{2}} (1-dm d\bar{m}) & dm dx_3 \end{array} \right)  \left( \begin{array}{c} \alpha \\ \beta \end{array} \right) 
\ee
To identify the primed spinor in a way that respects the complex conjugation
\be\label{ident-pr-even}
\left( \begin{array}{cc}  \frac{1}{\sqrt{2}} (1+dm d\bar{m}) & d\bar{m} dx_3 \end{array} \right)  \left( \begin{array}{c} \alpha^* \\ \beta^* \end{array} \right) 
\ee
we need to interpret the action of the $2\times 2$ matrices on spinors from the right instead of from the left. In this case the identification (\ref{ident-pr-even}) is compatible with the action (\ref{action-pr-even}). This explicitly verifies the expected fact that the Weyl representation of ${\rm SO}(3,3)$ splits into two Weyl representations of different chiralities under the action of ${\rm SO}(3,1)\subset{\rm SO}(3,3)$. 

Let us also describe the odd forms. We have the states that transform non-trivially under $K_3+\im P_3$
\be\label{primed*}
dm \qquad {\rm eigenvalue\,\, of\,\, } K_3 +\im P_3 \quad +\im, \\ \nonumber
\frac{1}{\sqrt{2}} (1-dm d\bar{m})dx_3  \qquad {\rm eigenvalue\,\, of \,\,} K_3 +\im P_3 \quad -\im,
\ee
as well as states that transform non-trivially under $K_3-\im P_3$
\be\label{unprimed*}
\frac{1}{\sqrt{2}} (1+dm d\bar{m})dx_3 \qquad {\rm eigenvalue\,\, of \,\,} K_3 -\im P_3 \quad +\im, \\ \nonumber
d\bar{m}  \qquad {\rm eigenvalue\,\, of \,\,} K_3 -\im P_3 \quad -\im.
\ee
These states are also eigenstates of $C$, with the primed spinor (\ref{primed*}) being of eigenvalue $-\im/2$ and the unprimed spinor (\ref{unprimed*}) of eigenvalue $+\im/2$. We thus get another complex 2-component spinor (and its complex conjugate) in the other Weyl representation of ${\rm SO}(3,3)$. 

If we now select the states of a given eigenvalue of $C$, e.g. $\im/2$, these are the unprimed spinor from the space of odd forms (\ref{unprimed*}) and the primed spinor from the space of even forms (\ref{primed}). Thus, the states of a given eigenvalue of $C$ form a charged Dirac spinor. Altogether, the Dirac representation of ${\rm SO}(3,3)$ is then a Dirac spinor of ${\rm SO}(3,1)$ and its complex conjugate spinor ${\bf 8}_\R={\bf 4}_\C$, where $\bf 4$ is the Dirac representation of the Lorentz group. 

For completeness, let us also list the action of operators (\ref{E-pm}), (\ref{bar-E-pm}) on the odd forms. We have
\be
E_- \frac{1}{\sqrt{2}} (1-dm d\bar{m})dx_3 = - \im dm, \qquad E_+ dm = - \im \frac{1}{\sqrt{2}} (1-dm d\bar{m})dx_3,
\ee
and for the unprimed spinor forms we have
\be
\bar{E}_- d\bar{m} = \im \frac{1}{\sqrt{2}} (1+ dm d\bar{m}) dx_3, \qquad 
\bar{E}_+ \frac{1}{\sqrt{2}} (1+dm d\bar{m}) dx_3 = \im d\bar{m}.
\ee
Comparison with (\ref{bar-E-2}) fixes the 2-component spinor identification up to an overall sign, which we choose as
\be\label{ident-un-odd}
\left( \begin{array}{cc} - d\bar{m} &  \frac{1}{\sqrt{2}} (1+dm d\bar{m}) dx_3 \end{array} \right) \left( \begin{array}{c} \alpha \\ \beta \end{array} \right) .
\ee
For the primed spinor, we again choose the identification that respects the complex conjugation
\be\label{ident-pr-odd}
\left( \begin{array}{cc} - dm &  \frac{1}{\sqrt{2}} (1-dm d\bar{m}) dx_3 \end{array} \right) \left( \begin{array}{c} \alpha^* \\ \beta^* \end{array} \right).
\ee

\subsection{The Dirac operator on $\R^{3,1}$}

Our goal now is to verify that the Dirac operator that arises from the exterior derivative on $\R^{3,3}$ reduces to the usual 4-dimensional Dirac operator when dimensionally reduced by assuming that all states depend only on $u^i, \tilde{u}^3$ but not on $\tilde{u}^1,\tilde{u}^2$. To this end, we need to describe the usual Dirac operator, analogously to what we have done for the Dirac operator on $\R^{2,2}$ in the previous section.

As the first step, we form the $2\times 2$ Hermitian matrix
\be\label{u}
{\bf u}^{A'A} =  \left( \begin{array}{cc} \tilde{u}^3+u^3 & u^1 - \im u^2 \\ u^1+\im u^2 & \tilde{u}^3-u^3  \end{array}\right).
\ee
This matrix has the property that its determinant is minus the interval
\be
-{\rm det}({\bf u}) = (u^1)^2 + (u^2)^2 +(u^3)^2 - (\tilde{u}^3)^2 \equiv |u|^2.
\ee
Form this we form the matrix with primed index down ${\bf u}_{A'}{}^A = -\epsilon_{A'B'} {\bf u}^{B'A}$ and so in index-free notations the matrix ${\bf u}_{A'}{}^A$ is given by minus $(\epsilon{\bf u})$, where ${\bf u}$ is the matrix (\ref{u}). We have
\be
{\bf u}_{A'}{}^A =  \left( \begin{array}{cc} -u^1-\im u^2  & u^3 - \tilde{u}^3  \\ u^3+\tilde{u}^3 & u^1 - \im u^2\end{array}\right).
\ee

We can also take the transpose of ${\bf u}$ and lower the unprimed index, i.e. consider $-\epsilon {\bf u}^T$. This gives
\be
{\bf u}_{A}{}^{A'} = \left( \begin{array}{cc} -u^1 + \im u^2   & u^3-\tilde{u}^3 \\ u^3 + \tilde{u}^3 &  u^1+\im u^2\end{array}\right).
\ee
Together, these matrices satisfy the properties
\be
{\bf u}_A{}^{B'} {\bf u}_{B'}{}^B = |u|^2 \id_A{}^B, \quad 
{\bf u}_{A'}{}^{B} {\bf u}_{B}{}^{B'} = |u|^2 \id_{A'}{}^{B'}.
\ee
We then define the two chiral Dirac operators
\be\label{chiral-dirac}
\partial_{A'}{}^A \equiv \partial = \left( \begin{array}{cc} -\partial/\partial u^1-\im \partial/\partial u^2 & \partial/\partial u^3 - \partial/\partial\tilde{u}^3 \\ \partial/\partial u^3+\partial/\partial\tilde{u}^3 & \partial/\partial u^1 - \im \partial/\partial u^2 \end{array}\right), \\ \nonumber
\partial_{A}{}^{A'} \equiv \partial^T = \left( \begin{array}{cc}-\partial/\partial u^1 + \im \partial/\partial u^2  & \partial/\partial u^3-\partial/\partial\tilde{u}^3 \\ \partial/\partial u^3 + \partial/\partial\tilde{u}^3 & \partial/\partial u^1+\im \partial/\partial u^2  \end{array}\right).
\ee

\subsection{Change of coordinates}

The derivatives appearing in the Dirac operator are with respect to $u^i$ and $\tilde{u}^3$, and these coordinates are related to the $x,\x$ coordinates as $u=(1/2)(x+\x), \tilde{u}=(1/2)(x-\x)$, and so we have
\be
\frac{\partial}{\partial u} = \frac{\partial}{\partial x} +\frac{\partial}{\partial \x} , \qquad \frac{\partial}{\partial \tilde{u}} = \frac{\partial}{\partial x} -\frac{\partial}{\partial \x},
\ee
where we omitted the indices for brevity. Thus, we can write all $u,\tilde{u}$ derivatives as combinations of $x,\x$ derivatives. 

Now, we introduce the complex linear combinations that appear in (\ref{dm-dbarm})
\be
m = \frac{1}{\sqrt{2}}( x_1 - \im x_2), \quad \bar{m} = \frac{1}{\sqrt{2}}( x_1 + \im x_2), \quad
\tilde{m} = \frac{1}{\sqrt{2}}( \x_1 - \im \x_2), \quad \tilde{\bar{m}} = \frac{1}{\sqrt{2}}( \x_1 + \im \x_2).
\ee
This gives
\be\label{xm-der-relations}
\frac{\partial }{\partial x_1} = \frac{1}{\sqrt{2}} \left( \frac{\partial}{\partial \bar{m}}+ \frac{\partial}{\partial m}  \right), \quad \frac{\partial }{\partial x_2} = \frac{\im}{\sqrt{2}} \left( \frac{\partial}{\partial \bar{m}} - \frac{\partial}{\partial m} \right), \\ \nonumber
\frac{\partial }{\partial \x_1} = \frac{1}{\sqrt{2}} \left( \frac{\partial}{\partial \tilde{\bar{m}}}+ \frac{\partial}{\partial \tilde{m}}  \right), \quad \frac{\partial }{\partial \x_2} = \frac{\im}{\sqrt{2}} \left( \frac{\partial}{\partial \tilde{\bar{m}}} - \frac{\partial}{\partial \tilde{m}} \right).
\ee
So, we can write all $u,\tilde{u}$ derivatives as combinations of derivatives with respect to $m,\bar{m},\tilde{m},\tilde{\bar{m}}$ and $x_3, \x_3$. 

We now take into account that the functions we want to consider, i.e. components of the differential forms, should depend only on $u^i, \tilde{u}^3$ but not on $\tilde{u}^1, \tilde{u}^2$. Because $\tilde{u}=(1/2)(x-\x)$, this means that all functions depend on $x_1, x_2, \x_1, \x_2$ only in combination $x_1 + \x_1, x_2+ \x_2$, but there is no dependence on the differences. This implies the following relations between the partial derivatives
\be\label{deriv-relations}
\frac{\partial}{\partial x_1} = \frac{\partial}{\partial \x_1}, \qquad
\frac{\partial}{\partial x_2} = \frac{\partial}{\partial \x_2}.
\ee
Then (\ref{xm-der-relations}) implies that 
\be\label{m-derivatives}
\frac{\partial}{\partial m} = \frac{\partial}{\partial \tilde{m}}, \qquad \frac{\partial}{\partial \bar{m}} =\frac{\partial}{\partial \tilde{\bar{m}}}.
\ee
These are the derivative relations that follow from the assumption of the dimensional reduction. Then the derivatives appearing in the Dirac operator are
\be
\frac{\partial}{\partial u^1} = 2 \frac{\partial }{\partial x_1} = \sqrt{2} \left( \frac{\partial}{\partial \bar{m}}+ \frac{\partial}{\partial m}  \right), \quad \frac{\partial}{\partial u^2} = 2 \frac{\partial }{\partial x_2} = \im \sqrt{2} \left( \frac{\partial}{\partial \bar{m}}- \frac{\partial}{\partial m}  \right),
\ee
and their complex linear combinations are
\be
\frac{\partial}{\partial u^1} + \im \frac{\partial}{\partial u^2}  = 2\sqrt{2} \frac{\partial }{\partial m}\equiv 2\sqrt{2} \partial_m, \qquad \frac{\partial}{\partial u^1} - \im \frac{\partial}{\partial u^2}  = 2\sqrt{2} \frac{\partial }{\partial \bar{m}}\equiv 2\sqrt{2} \partial_{\bar{m}}.
\ee
We also have the linear combinations of derivatives with respect to $u^3, \tilde{u}^3$
\be
\frac{\partial}{\partial u^3} + \frac{\partial}{\partial \tilde{u}^3} = 2\frac{\partial }{\partial x_3}\equiv 2\partial_3, \qquad \frac{\partial}{\partial u^3} - \frac{\partial}{\partial \tilde{u}^3} = 2\frac{\partial }{\partial \x_3}\equiv 2\tilde{\partial}_3.
\ee

Taking into account the above, we get the following expressions for the chiral Dirac operators (\ref{chiral-dirac})
\be\label{chiral-dirac*}
\partial = 2 \left( \begin{array}{cc} - \sqrt{2} \partial_m & \tilde{\partial}_3 \\ \partial_3 & \sqrt{2} \partial_{\bar{m}} \end{array} \right), \qquad \partial^T = 2 \left( \begin{array}{cc}-\sqrt{2} \partial_{\bar{m}}  & \tilde{\partial}_3 \\  \partial_3 & \sqrt{2} \partial_m \end{array} \right).
\ee
Note that $\partial^T = \overline{\partial}$, where bar denotes the complex conjugation. 

\subsection{The Lagrangian}

For future use, we now list the usual Lagrangian for a single Weyl 2-component fermion 
\be\label{L-weyl}
L= \im (\xi^\dagger)^{A'} \partial_{A'}{}^A \xi_A.
\ee
In index-free notations this reads
\be
L = -\im \xi^\dagger \epsilon \partial \xi = \im \left( \begin{array}{cc} \beta^* & - \alpha^* \end{array}\right)  \partial \left( \begin{array}{c} \alpha \\ \beta \end{array}\right).
\ee
The resulting action is explicitly real by integration by parts. This of course also follows directly from (\ref{L-weyl}) by Hermiticity of $\partial^{A'A}$. Neglecting the possible boundary terms we can also write the Lagrangian in an explicitly real form
\be\label{L-real}
L = -\frac{\im}{2} \xi^\dagger \epsilon \partial  \xi + {\rm c.c.} 
\ee

\subsection{The Dirac operator as the exterior derivative}

Using (\ref{m-derivatives}), the $D$ operator applied to the differential form corresponding to a primed spinor is
\be\nonumber
D\left( \alpha^* \frac{1}{\sqrt{2}} (1+d m d\bar{m}) + \beta^* d\bar{m} dx_3\right) = \frac{1}{\sqrt{2}}\left( \partial_m \alpha^* dm + \partial_{\bar{m}} \alpha^* d\bar{m} + \partial_m \alpha^* d\tilde{m} + \partial_{\bar{m}} \alpha^* d\tilde{\bar{m}}\right) (1+dm d\bar{m}) \\ \nonumber
+ \frac{1}{\sqrt{2}} \partial_3 \alpha^* dx_3 (1+dm d\bar{m}) + \partial_m \beta^* dm d\bar{m} dx_3 + (\partial_m \beta^* d\tilde{m} + \tilde{\partial}_3 \beta^* d\tilde{x}_3) \bar{m} dx_3,
\ee
where we only wrote non-vanishing terms. The using the rule that $d\tilde{m}$ eats $d\bar{m}$ and $d\tilde{\bar{m}}$ eats $dm$ we see that there are some cancellations in the above expression, and it simplifies to
\be
-(\tilde{\partial}_3 \beta^* - \sqrt{2} \partial_{\bar{m}} \alpha^* ) d\bar{m} +( \partial_3 \alpha^* + \sqrt{2} \partial_m \beta^*) \frac{1}{\sqrt{2}}(1+dm d\bar{m}) dx_3. 
\ee
This result can be written as
\be\label{D-primed}
D\left( \left( \begin{array}{cc} \frac{1}{\sqrt{2}} (1+dm d\bar{m}) & d\bar{m} dx_3 \end{array}\right) \left( \begin{array}{c} \alpha^* \\ \beta^* \end{array}\right) \right) =  \left( \begin{array}{cc} - d\bar{m} & \frac{1}{\sqrt{2}} (1+dm d\bar{m})  dx_3 \end{array}\right) \frac{1}{2} \partial^T \left( \begin{array}{c} \alpha^* \\ \beta^* \end{array}\right) ,
\ee
where we have used (\ref{chiral-dirac*}). The comparison with (\ref{ident-un-odd}) shows that $D$ gives the correction action of the Dirac operator, up to a factor of $1/2$. 

We now do a similar calculation for the unprimed spinor form, skipping intermediate steps this time. We have
\be
D\left( \alpha \frac{1}{\sqrt{2}} (1-dm d\bar{m}) + \beta dm dx_3 \right) \\ \nonumber
= ( \partial_3 \alpha + \sqrt{2} \partial_{\bar{m}} \beta) \frac{1}{\sqrt{2}} (1-dm d\bar{m}) dx_3 - (\tilde{\partial}_3 \beta - \sqrt{2} \partial_m \alpha) dm.
\ee
This can be written as follows
\be\label{D-unprimed}
D\left( \left( \begin{array}{cc} \frac{1}{\sqrt{2}} (1-dm d\bar{m}) & dm dx_3 \end{array}\right) \left( \begin{array}{c} \alpha \\ \beta \end{array}\right) \right) =  \left( \begin{array}{cc} - dm & \frac{1}{\sqrt{2}} (1-dm d\bar{m})  dx_3 \end{array}\right) \frac{1}{2} \partial \left( \begin{array}{c} \alpha \\ \beta \end{array} \right).
\ee
Again, comparison with (\ref{ident-pr-odd}) shows that the correction action of the chiral Dirac operator is reproduced. This formula is just the complex conjugate of (\ref{D-primed}) as it should be. 

It can also be similarly verified that
\be\label{D-unprimed'}
D\left( \left( \begin{array}{cc} -d\bar{m}& \frac{1}{\sqrt{2}} (1+dm d\bar{m})dx_3 \end{array}\right) \left( \begin{array}{c} \alpha \\ \beta \end{array}\right) \right) = 
\left( \begin{array}{cc} \frac{1}{\sqrt{2}} (1+dm d\bar{m})  & d\bar{m} dx_3   \end{array}\right) \frac{1}{2}\partial \left( \begin{array}{c} \alpha \\ \beta \end{array} \right),
\ee
and that the complex conjugate version of this formula also holds. Overall, the operator $D$ reproduces the chiral Dirac operators. 

\subsection{The Lagrangian in terms of differential forms}

We take a real Weyl spinor of ${\rm SO}(3,3)$ which is realised in the space of even forms in $\R^3$. In terms of introduced above basis (\ref{ident-un-even}), (\ref{ident-pr-even}) this corresponds to the following differential form
\be
\Psi = \left( \begin{array}{cc} \frac{1}{\sqrt{2}} (1-dm d\bar{m}) & dm dx_3 \end{array}\right) \left( \begin{array}{c} \alpha \\ \beta \end{array}\right) + {\rm c.c.},
\ee
where c.c. stands for the complex conjugate form. To simplify calculations that follow, it is convenient to rewrite this state in index-free notations as
\be\label{f-even}
\Psi = f_{even}^T \xi + \xi^\dagger f_{even}^*, \qquad f_{even}:= \left( \begin{array}{c} \frac{1}{\sqrt{2}} (1-dm d\bar{m}) \\ dm dx_3 \end{array}\right) , \qquad \xi:= \left( \begin{array}{c} \alpha \\ \beta \end{array}\right),
\ee
where we introduced a column of even forms that correspond to an unprimed Lorentz spinor. 
The $D$ derivative of $f_{even}^T \xi $ is given by (\ref{D-unprimed}). We can rewrite this derivative as
\be
D( f_{even}^T \xi) = f^\dagger_{odd} \frac{1}{2} \partial \xi,
\ee
where we introduced 
\be\label{f-odd}
f_{odd} = \left( \begin{array}{c} -d\bar{m}\\ \frac{1}{\sqrt{2}} (1+dm d\bar{m})dx_3 \end{array}\right) .
\ee

We now calculate $(\Psi,D\Psi)$. We will use the first inner product described in (\ref{inner}). We need the following matrix of inner products 
\be
(f^*_{even}, f^\dagger_{odd})=\left( \left( \begin{array}{c} \frac{1}{\sqrt{2}} (1+dm d\bar{m}) \\ d\bar{m} dx_3   \end{array}\right),\left( \begin{array}{cc}   -dm  & \frac{1}{\sqrt{2}} (1-dm d\bar{m})dx_3  \end{array}\right)  \right)\\ \nonumber
= \left( \begin{array}{c} \frac{1}{\sqrt{2}} (1-dm d\bar{m}) \\ -d\bar{m} dx_3   \end{array}\right) \left( \begin{array}{cc}   -dm  & \frac{1}{\sqrt{2}} (1-dm d\bar{m})dx_3  \end{array}\right) \Big|\\ \nonumber
= -\epsilon dm d\bar{m} dx_3 \Big| = -\im \epsilon,
\ee
where $\epsilon$ is the anti-symmetric matrix (\ref{epsilon}), and we have used $dm d\bar{m} = \im dx_1 dx_2$.

This means that
\be
(\Psi, D\Psi) = \xi^\dagger (-\im \epsilon) \frac{1}{2} \partial \xi + {\rm c.c.} =L,
\ee
where we have compared with (\ref{L-real}). Thus, the Lagrangian  in terms of differential forms $(\Psi, D\Psi)$, which is explicitly real, matches the Weyl Lagrangian (\ref{L-weyl}) in its explicitly real form (\ref{L-real}).

\section{The case of ${\rm SO}(5,5)$}
\label{sec:55}

This case is interesting because the stabiliser subgroup of the Lorentz group in this case is ${\rm SO}(2,4)$, the conformal group in 4 dimensions. In particular, the maximally compact subgroup of the stabiliser is ${\rm SO}(2)\sim {\rm U}(1)$ times ${\rm SO}(4)={\rm SU}(2)\times{\rm SU}(2)/\Z_2$. So, our Lorenz spinors will also transform non-trivially under ${\rm U}(1)\times {\rm SU}(2)\times{\rm SU}(2)$, which is similar to what is happening in the SM, and so is interesting. However, as we know from general considerations, in this case the Weyl Lagrangian $(\Psi, D\Psi)$ vanishes for either of the two inner products. Thus, there is no interesting Lagrangian that can be written in this case. Still, we work it out because there are many similarities with the representation theory of the interesting ${\rm SO}(7,7)$ case. 

\subsection{Embedding of ${\rm SO}(3,1)\times{\rm SO}(2)\times{\rm SO}(4)$}

We now have two more coordinate pairs $u^{4,5},\tilde{u}^{4,5}$ and $x_{4,5}, \x_{4,5}$ as compared to the previous section. We leave the embedding of the Lorentz Lie algebra unchanged, and given by (\ref{Lorentz-K}), (\ref{Lorentz-P}). So, we only need to describe the embedding of ${\rm SO}(2)\times{\rm SO}(4)$.

Let us start with ${\rm SO}(2)$. This group mixes the coordinates $u^4, u^5$. Its Lie algebra in the $x,\x$ basis is then given by matrices of the type (\ref{rotations}), where $\beta_{ij}= \delta_{[i4} \delta_{j]5}$. The corresponding operator, which we shall call $C$ (for charge), is given by
\be
C = -\frac{1}{2}\left( a_4 a_5^\dagger - a_5 a_4^\dagger + a_4 a_5 + a_4^\dagger a_5^\dagger\right).
\ee

Let us now describe the ${\rm SO}(4)$ subgroup. This is the group of rotations that mixes the coordinates $\tilde{u}^{1,2,4,5}$. In the $x,\x$ basis it consists of matrices of the form (\ref{rotations-dual}). We will call the generators $\bK_i,\bP_i$ and choose them as follows
\be
\bK_1 = -\frac{1}{2} \left( a_2 a_4^\dagger - a_4 a_2^\dagger - a_2 a_4 - a_2^\dagger a_4^\dagger\right), \quad \bK_2 = -\frac{1}{2} \left( a_4 a_1^\dagger - a_1 a_4^\dagger - a_4 a_1 - a_4^\dagger a_1^\dagger\right), \\ \nonumber
\bK_3 = -\frac{1}{2} \left( a_1 a_2^\dagger - a_1 a_2^\dagger - a_1 a_2 - a_1^\dagger a_2^\dagger\right),
\ee
and
\be
\bP_1 = -\frac{1}{2} \left( a_1 a_5^\dagger - a_5 a_1^\dagger - a_1 a_5 - a_1^\dagger a_5^\dagger\right), \quad \bP_2 = -\frac{1}{2} \left( a_2 a_5^\dagger - a_5 a_2^\dagger - a_2 a_5 - a_2^\dagger a_5^\dagger\right), \\ \nonumber
\bP_3 = -\frac{1}{2} \left( a_4 a_5^\dagger - a_5 a_4^\dagger - a_4 a_5 - a_4^\dagger a_5^\dagger\right).
\ee
They generate the ${\rm SO}(4)$ Lie algebra
\be
[ \bK_i, \bK_j]=\epsilon_{ij}{}^k \bK_k, \qquad [\bK_i,\bP_j]=\epsilon_{ij}{}^k \bP_k, \qquad [\bP_i,\bP_j]=\epsilon_{ij}{}^k \bK_k.
\ee

\subsection{Change of basis}

We now go to the $m,\bar{m}$ basis (\ref{dm-dbarm}). In this basis the operator $\bK_3$ takes the following form
\be
\bK_3 = -\frac{\im}{2} \left( a_{\bar{m}} a_m^\dagger - a_m a_{\bar{m}}^\dagger - a_{\bar{m}} a_m - a_{\bar{m}}^\dagger a_m^\dagger \right).
\ee
This immediately gives the eingestates of $\bK_3$
\be\label{bK3-action}
\bK_3 dm = \frac{\im}{2} dm, \quad \bK_3 d\bar{m} = -\frac{\im}{2}d\bar{m} , \quad 
\bK_3 (1\pm dm d\bar{m}) = \mp \frac{\im}{2} (1\pm dm d\bar{m}).
\ee 

To describe the other operators, let us introduce a set of complex coordinates $z,\bar{z}$ mixing the directions $4,5$
\be\label{dz-dbarz}
dz=\frac{1}{\sqrt{2}}( dx_4 - \im dx_5), \qquad d\bar{z}=\frac{1}{\sqrt{2}}( dx_4 + \im dx_5).
\ee
This is in complete parallel with (\ref{dm-dbarm}). We then define a new set of creation and annihilation operators, corresponding to creation-annihilation of $z,\bar{z}$
\be\label{az-zbar}
a_z := \frac{1}{\sqrt{2}}( a_4 -\im a_5), \qquad a_{\bar{z}} := \frac{1}{\sqrt{2}}( a_4 +\im a_5), \\ \nonumber
a_z^\dagger := \frac{1}{\sqrt{2}}( a_4^\dagger -\im a_5^\dagger), \qquad a_{\bar{z}}^\dagger := \frac{1}{\sqrt{2}}( a_4^\dagger +\im a_5^\dagger).
\ee
The non-trivial anti-commutation relations are 
\be
a_z a_{\bar{z}}^\dagger + a_{\bar{z}}^\dagger a_z = 1, \qquad a_{\bar{z}} a_z^\dagger + a_z^\dagger a_{\bar{z}} = 1.
\ee

In terms of the new basis, the operator $\bP_3$ takes the form
\be
\bP_3 = -\frac{\im}{2} \left( a_{\bar{z}} a_z^\dagger - a_z a_{\bar{z}}^\dagger - a_{\bar{z}} a_z - a_{\bar{z}}^\dagger a_z^\dagger \right).
\ee
The eingestates of $\bK_3$ are
\be\label{bP3-action}
\bP_3 dz = \frac{\im}{2} dz, \quad \bP_3 d\bar{z} = -\frac{\im}{2}d\bar{z} , \quad 
\bP_3 (1\pm dz d\bar{z}) = \mp \frac{\im}{2} (1\pm dz d\bar{z}).
\ee 
We can also write down the $C$ operator in the new basis
\be
C = -\frac{\im}{2} \left( a_{\bar{z}} a_z^\dagger - a_z a_{\bar{z}}^\dagger + a_{\bar{z}} a_z + a_{\bar{z}}^\dagger a_z^\dagger \right),
\ee
with eigenstates
\be\label{bP3-action}
C dz = \frac{\im}{2} dz, \quad C d\bar{z} = -\frac{\im}{2}d\bar{z} , \quad 
C (1\pm dz d\bar{z}) = \pm \frac{\im}{2} (1\pm dz d\bar{z}).
\ee 

Let us also spell out the complex linear combinations operators. We have
\be\label{bK-pm}
\frac{1}{\sqrt{2}}(\bK_1 -\im \bK_2) = -\frac{\im}{2} ( a_m a_4^\dagger - a_4 a_m^\dagger - a_m a_4 - a_m^\dagger a_4^\dagger ), \\ \nonumber
\frac{1}{\sqrt{2}}(\bK_1 +\im \bK_2) = \frac{\im}{2} ( a_{\bar{m}} a_4^\dagger - a_4 a_{\bar{m}}^\dagger - a_{\bar{m}} a_4 - a_{\bar{m}}^\dagger a_4^\dagger ),
\ee
and
\be\label{bP-pm}
\frac{1}{\sqrt{2}}(\bP_1 -\im \bP_2) = -\frac{1}{2} ( a_m a_5^\dagger - a_5 a_m^\dagger - a_m a_5 - a_m^\dagger a_5^\dagger ), \\ \nonumber
\frac{1}{\sqrt{2}}(\bP_1 +\im \bP_2) = -\frac{1}{2} ( a_{\bar{m}} a_5^\dagger - a_5 a_{\bar{m}}^\dagger - a_{\bar{m}} a_5 - a_{\bar{m}}^\dagger a_5^\dagger ).
\ee

Finally, we introduce the self-dual/anti-self-dual combinations
\be\label{F-pm}
F_-:=\frac{1}{2}(\bK_1-\im \bK_2) +\frac{1}{2}(\bP_1-\im \bP_2) = -\frac{\im}{2} \left( a_m a_z^\dagger - a_z a_m^\dagger - a_m a_z - a_m^\dagger a_z^\dagger   \right), \\ \nonumber
F_+:=\frac{1}{2}(\bK_1+\im \bK_2) +\frac{1}{2} (\bP_1+\im \bP_2) = \frac{\im}{2} \left( a_{\bar{m}} a_{\bar{z}}^\dagger - a_{\bar{z}} a_{\bar{m}}^\dagger - a_{\bar{m}} a_{\bar{z}} - a_{\bar{m}}^\dagger a_{\bar{z}}^\dagger   \right),
\ee
and
\be\label{bF-pm}
\bar{F}_-:=\frac{1}{2}(\bK_1-\im \bK_2) -\frac{1}{2} (\bP_1-\im \bP_2) = -\frac{\im}{2} \left( a_m a_{\bar{z}}^\dagger - a_{\bar{z}} a_m^\dagger - a_m a_{\bar{z}} - a_m^\dagger a_{\bar{z}}^\dagger   \right), \\ \nonumber
\bar{F}_+:=\frac{1}{2}(\bK_1+\im \bK_2) -\frac{1}{2} (\bP_1+\im \bP_2) = \frac{\im}{2} \left( a_{\bar{m}} a_z^\dagger - a_z a_{\bar{m}}^\dagger - a_{\bar{m}} a_z - a_{\bar{m}}^\dagger a_z^\dagger   \right).
\ee

\subsection{Eigenstates}

We now consider the space of even forms on $\R^5$. In comparison to the situation analysed in the previous section, we have added two new coordinates, which in the complex basis are $z,\bar{z}$. Thus, the even forms on $\R^5$ are even forms on $\R^3$ that we already know how to interpret times even forms on $\R^2$, of which the most convenient basis is $(1\pm dz d\bar{z})$, plus odd forms on $\R^3$ times odd forms on $\R^2$ for which the basis is $dz, d\bar{z}$. We should now construct combinations that, in addition to diagonalising operators $K_3\pm \im P_3$, also diagonalise $\bK_3 \pm \bP_3$, as well as $C$. At this stage it is best to pass to particle physics notations, and label states by the eigenstates of all these operators. 

Let us first work out the eigenstates of $\bK_3\pm \bP_3$. All eigenvalues will be $\pm \im$, and for the copy of ${\rm SU}(2)$ to which the generator $\bK_3 +\bP_3$ belongs (we will refer to this copy as right) we will identify the eigenvalue $+\im$ as projection of spin $+(1/2)$, and eigenvalue $-\im$ as projection of spin $-(1/2)$. For the copy of ${\rm SU}(2)$ containing the generator $\bK_3 -\bP_3$ (we will refer to it as left) the identification is opposite. When listing the eigenvalues, we will list the left eigenvalue first, followed by the right eigenvalue. We will also indicate the eigenvalue of $C$, in the third position, with eigenvalue $+\im/2$ indicated as $+1$ and $-\im/2$ as $-1$.

This conventions produce the following list of states. We start with the eigenstates of ${\rm SU}(2)_R$
\be
dm dz \qquad \left( 0, \frac{1}{2},+1 \right) \qquad \frac{1}{2}(1+dm d\bar{m})(1+ dz d\bar{z}) \qquad \left( 0, -\frac{1}{2},+1 \right), \\ \nonumber
\frac{1}{2}(1-dm d\bar{m})(1- dz d\bar{z}) \qquad \left( 0, \frac{1}{2},-1\right) \qquad d\bar{m} d\bar{z} \qquad \left( 0, -\frac{1}{2},-1\right) 
\ee
in the space of even forms in $\R^4$ and
\be
\frac{1}{\sqrt{2}} dm (1- dz d\bar{z}) \qquad \left( 0, \frac{1}{2},-1\right) \qquad \frac{1}{\sqrt{2}} d\bar{z} (1 +dm d\bar{m}) \qquad \left( 0, -\frac{1}{2},-1\right) \\ \nonumber
\frac{1}{\sqrt{2}} dz (1 -dm d\bar{m})\qquad \left( 0, \frac{1}{2},+1\right) \qquad \frac{1}{\sqrt{2}} d\bar{m} (1+ dz d\bar{z})\qquad \left( 0, -\frac{1}{2},+1\right)
\ee
in the space of odd forms. Note that the complex conjugate of a right doublet is a right doublet, as it should be. Altogether we have 4 ${\rm SU}(2)_R$ doublets, 2 doublets of each charge. The ${\rm SU}(2)_L$ doublets in the space of even forms in $\R^4$ are
\be
d\bar{m} dz \qquad \left( \frac{1}{2}, 0,+1\right) \qquad \frac{1}{2}(1-dm d\bar{m})(1+ dz d\bar{z}) \qquad \left( -\frac{1}{2},0,+1\right), \\ \nonumber
\frac{1}{2}(1+dm d\bar{m})(1- dz d\bar{z}) \qquad \left( \frac{1}{2}, 0,-1\right) \qquad dm d\bar{z} \qquad \left( -\frac{1}{2},0,-1\right) ,
\ee
and in the space of odd forms
\be
\frac{1}{\sqrt{2}} d\bar{m} (1- dz d\bar{z}) \qquad \left( \frac{1}{2},0,-1\right) \qquad \frac{1}{\sqrt{2}} d\bar{z} (1 -dm d\bar{m}) \qquad \left( -\frac{1}{2},0,-1\right) \\ \nonumber
\frac{1}{\sqrt{2}} dz (1 +dm d\bar{m})\qquad \left( \frac{1}{2},0,+1\right) \qquad \frac{1}{\sqrt{2}} dm (1+ dz d\bar{z})\qquad \left(-\frac{1}{2},0,+1\right).
\ee
Again this gives 4 ${\rm SU}(2)_L$ doublets.

Finally, by wedging the odd forms from the above list with $dx_3$ we get all of 16 even forms in $\R^5$. They are also eigenstates of $K_3\pm \im P_3$, as we know from the previous section. Let us select from this list of 16 states 8 states that are left (unprimed) spinors, i.e. transform non-trivially under $K_3-\im P_3$. The required states can be seen in (\ref{ident-un-even}), (\ref{ident-un-odd}).

Another useful piece of notation is to introduce particle names for the listed above states. Thus, we shall refer to eigenstates of ${\rm SU}(2)_L$ using unbarred letters, and will put a bar above the name of a particle for ${\rm SU}(2)_R$ states. This gives
\be
\left( \begin{array}{c} \nu \\ e \end{array}\right) = \left( \begin{array}{c} +1/2 \\ -1/2 \end{array}\right)_L, \qquad \left( \begin{array}{c} \bar{\nu} \\ \bar{e} \end{array}\right) = \left( \begin{array}{c} +1/2 \\ -1/2 \end{array}\right)_R.
\ee
Here the bars are just parts of the name and have nothing to do with complex conjugation. With these notations in mind we can write the form that corresponds to an unprimed Lorentz spinor and an ${\rm SU}(2)_L$ spinor. We have
\be
\Psi_L := \left( \begin{array}{cc} -d\bar{m} dz& \frac{1}{\sqrt{2}} (1+dm d\bar{m})dx_3 dz\end{array}\right) \left( \begin{array}{c} \alpha_\nu \\ \beta_\nu \end{array}\right) 
\\ \nonumber
+\left( \begin{array}{cc} \frac{1}{2} (1-dm d\bar{m})(1+dz d\bar{z})  & \frac{1}{\sqrt{2}} dm dx_3 (1+dz d\bar{z}) \end{array}\right) \left( \begin{array}{c} \alpha_e \\ \beta_e \end{array}\right) ,
\ee
where the signs are chosen as in the previous section. This left spinor is of $C$ charge $+1$ in the conventions chosen. We can similarly write the form that corresponds to an unprimed Lorentz spinor and an ${\rm SU}(2)_R$ spinor. We have
\be
\Psi_R:= \left( \begin{array}{cc} \frac{1}{2} (1-dm d\bar{m})(1-dz d\bar{z})  & \frac{1}{\sqrt{2}} dm dx_3 (1-dz d\bar{z}) \end{array}\right) \left( \begin{array}{c} \alpha_{\bar{\nu}} \\ \beta_{\bar{\nu}} \end{array}\right)\\ \nonumber
+\left( \begin{array}{cc} -d\bar{m} d\bar{z}& \frac{1}{\sqrt{2}} (1+dm d\bar{m})dx_3 d\bar{z}\end{array}\right) \left( \begin{array}{c} \alpha_{\bar{e}} \\ \beta_{\bar{e}} \end{array}\right) .
\ee 
The right spinor is of $C$ charge $-1$. This parametrises the general real even form on $\R^5$ as 
\be
\Psi := \Psi_L + \Psi_R + {\rm c.c.}
\ee

\subsection{Rewriting}

We now rewrite the above form in a way more suitable for computations. We have already introduced the Lorentz 2-component states (\ref{f-even}) and (\ref{f-odd}). With these notations we can rewrite the form $\Psi_L$ as
\be
\Psi_L = f^T_{\rm odd} dz \nu + f^T_{even} \frac{1}{\sqrt{2}} (1+ dz d\bar{z}) e,
\ee
where it is understood that $\nu, e$ are 2-component unprimed Lorentz spinors
\be
\nu = \left( \begin{array}{c} \alpha_{\bar{\nu}} \\ \beta_{\bar{\nu}} \end{array}\right), \qquad e = \left( \begin{array}{c} \alpha_{\bar{e}} \\ \beta_{\bar{e}} \end{array}\right).
\ee
Similarly,
\be
\Psi_R = f^T_{even} \frac{1}{\sqrt{2}} (1- dz d\bar{z}) \bar{\nu} + f^T_{\rm odd} d\bar{z} \bar{e}.
\ee

\subsection{The Lagrangian}

We know from general considerations that the Lagrangian $(\Psi,D\Psi)$ should be zero in this case, modulo surface terms. Here we confirm this by an explicit calculation, to become more proficient with the technology.

With results (\ref{D-unprimed}), (\ref{D-unprimed'}) it is easy to compute the action of the $D$ operator. Indeed, the component functions have dependence on only $m,\bar{m},x_3$ and $\tilde{m},\tilde{\bar{m}}, \x_3$, and there is no dependence on the $z$-coordinates. Thus, the $D$ operator is insensitive to the terms that contain $dz, d\bar{z}$. These terms are always written at the end of the forms, and they continue to remain there after the $D$ acts on what stands before such terms according to (\ref{D-unprimed}), (\ref{D-unprimed'}). Thus, we can use
\be\label{D-action}
D(f^T_{even} \xi) = f^\dagger_{odd} \frac{1}{2} \partial \xi, \qquad D(f^T_{odd} \xi) = f^\dagger_{even} \frac{1}{2} \partial \xi.
\ee
This gives
\be
D\Psi_L = f^\dagger_{even} dz \frac{1}{2}\partial \nu + f^\dagger_{odd} \frac{1}{\sqrt{2}} (1+ dz d\bar{z})   \frac{1}{2} \partial e ,
\ee
and
\be
D\Psi_R = f^\dagger_{odd} \frac{1}{\sqrt{2}} (1- dz d\bar{z}) \frac{1}{2} \partial \bar{\nu}  
+ f^\dagger_{even} d\bar{z} \frac{1}{2}\partial \bar{e}.
\ee

We now need to compute the relevant inner products of forms. An example calculation is
\be
( f^*_{odd} d\bar{z} , f^\dagger_{even} dz) = \left( \left( \begin{array}{c}   -dm d\bar{z} \\ \frac{1}{\sqrt{2}} (1-dm d\bar{m})dx_3 d\bar{z} \end{array}\right), \left( \begin{array}{cc} \frac{1}{\sqrt{2}} (1+dm d\bar{m})dz  & d\bar{m} dx_3 dz  \end{array}\right) \right)\\ \nonumber
= \left( \begin{array}{c}   dm d\bar{z} \\ \frac{1}{\sqrt{2}} (-1-dm d\bar{m})dx_3 d\bar{z} \end{array}\right)\wedge \left( \begin{array}{cc} \frac{1}{\sqrt{2}} (1+dm d\bar{m})dz  & d\bar{m} dx_3 dz  \end{array}\right) \Big|\\ \nonumber
= -\epsilon dm d\bar{m} dx_3 dz d\bar{z} \Big| = -(\im)^2 \epsilon = \epsilon,
\ee
where $\epsilon$ is the anti-symmetric matrix (\ref{epsilon}). We note that we could have computed this in steps using $(f^*_{odd}, f^\dagger_{even})=-\im \epsilon$. Indeed
\be
( f^*_{odd} d\bar{z} , f^\dagger_{even} dz) = - (d\bar{z} f^*_{odd} , f^\dagger_{even} dz)= -(d\bar{z}, dz) (f^*_{odd}, f^\dagger_{even}) = \im (-\im) \epsilon= \epsilon,
\ee
where we used $-(d\bar{z}, dz) = - d\bar{z} dz \Big|= dz d\bar{z} \Big| = \im$. 

This means that the part of the Lagrangian $(\Psi, D\Psi)$ that depends on the $\nu$ functions is
\be\label{nu-terms}
\nu^\dagger \epsilon \frac{1}{2} \partial \nu + {\rm c.c.} = (\nu^\dagger)_{A'} \partial^{A'A} \nu_A + {\rm c.c.}
\ee
We should notice however that there is no factor of $\im$ in this Lagrangian, in contrast to what happened in the ${\rm SO}(3,3)$ setup. This happened here because we got two factors of $\im$, one coming from $dm d\bar{m}$, and another coming from $dz d\bar{z}$. But because there is no factor of $\im$ here we get a cancellation, modulo surface terms. Indeed, because $\partial^{A'A}$ is a Hermitian matrix we have, for any spinor $\xi_A$
\be
\left(\int  (\xi^\dagger)_{A'} \partial^{A' A}\xi_A\right)^\dagger =  \int (\partial^{A' A}\xi_A)^\dagger \xi_A =\int \partial^{A' A}(\xi^\dagger)_{A'} \xi_A = -\int (\xi^\dagger)_{A'} \partial^{A' A} \xi_A,
\ee
and so the complex conjugate of the first term in (\ref{nu-terms}) is minus itself modulo surface terms, which is the reason why in (\ref{L-weyl}) we multiply this expression by $\im$ to get a real Lagrangian. But this means that there is cancellation in (\ref{nu-terms}). Thus, we confirmed by an explicit verification that $(\Psi, D\Psi)=0$ modulo surface terms in the ${\rm SO}(5,5)$ setup. Thus, there is no interesting Lagrangian we can write in this case. 

 \section{The case of ${\rm SO}(7,7)$}
 \label{sec:77}
 
 We now come to the case of real interest. We embed the Lorentz group ${\rm SO}(3,1)$ as described before. Its commutant in ${\rm SO}(7,7)$ is ${\rm SO}(4,6)$, and its compact subgroup is ${\rm SO}(4)\times {\rm SO}(6)$, which is the Pati-Salam GUT gauge group. We want to describe how even forms in $\R^7$ give the desired SM fermion representations. 
 
 \subsection{The group ${\rm SO}(4)$}
 
 Let us start by describing the group that rotates the $4,5,6,7$ directions. In the diagonal $u,\tilde{u}$ basis it is represented by matrices occupying the upper left corner. As we have already deduced when considering $1,2,3$ rotations, in the $x,\x$ basis this corresponds to matrices of the form (\ref{rotations}). In terms of creation-annihilation operators we get the following set
 \be
 \bK_1 = -\frac{1}{2} \left( a_5 a_6^\dagger - a_6 a_5^\dagger + a_5 a_6 + a_5^\dagger a_6^\dagger \right) , \quad  \bK_2 = -\frac{1}{2} \left( a_6 a_4^\dagger - a_4 a_6^\dagger + a_6 a_4 + a_6^\dagger a_4^\dagger \right), \\ \nonumber
  \bK_3 = -\frac{1}{2} \left( a_4 a_5^\dagger - a_5 a_4^\dagger + a_4 a_5 + a_4^\dagger a_5^\dagger \right),
 \ee
 and
 \be
 \bP_1 = -\frac{1}{2} \left( a_7 a_4^\dagger - a_4 a_7^\dagger + a_7 a_4 + a_7^\dagger a_4^\dagger \right) , \quad  \bP_2 = -\frac{1}{2} \left( a_7 a_5^\dagger - a_5 a_7^\dagger + a_7 a_5 + a_7^\dagger a_5^\dagger \right), \\ \nonumber
  \bP_3 = -\frac{1}{2} \left( a_7 a_6^\dagger - a_6 a_7^\dagger + a_7 a_6 + a_7^\dagger a_6^\dagger \right).
 \ee
 The commutators between these read
 \be
[ \bK_i,\bK_j]= \epsilon_{ijk}  \bK_k, \quad [  \bK_i, \bP_j]= \epsilon_{ijk}  \bP_k, \quad
[  \bP_i, \bP_j]= \epsilon_{ijk}  \bK_k.
\ee

\subsection{The group ${\rm SO}(6)$}

We will only describe explicitly the maximally commuting set of generators of ${\rm SO}(6)\sim {\rm SU}(4)$. The group ${\rm SO}(6)$ is the one mixing directions $\tilde{u}^{1,2}, \tilde{u}^{4,5,6,7}$. It thus consists of matrices of the form (\ref{rotations-dual}). So, we take the following maximally commuting set of generators 
\be\label{T}
T_1 =  -\frac{1}{2} \left( a_1 a_2^\dagger - a_1 a_2^\dagger - a_1 a_2 - a_1^\dagger a_2^\dagger\right), T_2 = -\frac{1}{2} \left( a_4 a_5^\dagger - a_5 a_4^\dagger - a_4 a_5 - a_4^\dagger a_5^\dagger\right), \\ \nonumber
T_3 = -\frac{1}{2} \left( a_6 a_7^\dagger - a_7 a_6^\dagger - a_6 a_7 - a_6^\dagger a_7^\dagger\right).
\ee

\subsection{Change of  basis}

We now go to the basis of coordinates $m,\bar{m}$ in the $x_1, \x_2$ space and $z,\bar{z}$ in $x_4, x_5$ space, and introduce 
\be
w=\frac{1}{\sqrt{2}}( x_6 - \im x_7), \qquad \bar{w}=\frac{1}{\sqrt{2}}( x_6 + \im x_7).
\ee
We then define a new set of creation and annihilation operators, corresponding to creation-annihilation of $dw,d\bar{w}$, with all formulas being analogous to (\ref{az-zbar}). 

In terms of the new creation-annihilation operators, the maximally commuting set of ${\rm SO}(4)$ generators takes the form
\be
\bK_3 = -\frac{\im}{2} \left( a_{\bar{z}} a_z^\dagger - a_z a_{\bar{z}}^\dagger + a_{\bar{z}} a_z + a_{\bar{z}}^\dagger a_z^\dagger \right), \bP_3 = \frac{\im}{2} \left( a_{\bar{w}} a_w^\dagger - a_w a_{\bar{w}}^\dagger + a_{\bar{w}} a_w + a_{\bar{w}}^\dagger a_w^\dagger \right).
\ee
On the other hand, the generators (\ref{T}) take the following form
\be
T_1 = -\frac{\im}{2} \left( a_{\bar{m}} a_m^\dagger - a_m a_{\bar{m}}^\dagger - a_{\bar{m}} a_m - a_{\bar{m}}^\dagger a_m^\dagger \right), \quad T_2 = -\frac{\im}{2} \left( a_{\bar{z}} a_z^\dagger - a_z a_{\bar{z}}^\dagger - a_{\bar{z}} a_z - a_{\bar{z}}^\dagger a_z^\dagger \right), \\ \nonumber
T_3 = -\frac{\im}{2} \left( a_{\bar{w}} a_w^\dagger - a_w a_{\bar{w}}^\dagger - a_{\bar{w}} a_w - a_{\bar{w}}^\dagger a_w^\dagger \right).
\ee

We have the following eigenstates of $\bK_3$
\be
\bK_3 dz = \frac{\im}{2} dz, \quad \bK_3 d\bar{z} = -\frac{\im}{2}d\bar{z} , \quad 
\bK_3(1\pm dz d\bar{z}) = \pm \frac{\im}{2} (1\pm dz d\bar{z}),
\ee 
and of $\bP_3$
\be
\bP_3 dw = -\frac{\im}{2} dw, \quad \bP_3 d\bar{w} = \frac{\im}{2}d\bar{w} , \quad 
\bP_3(1\pm dw d\bar{w}) = \mp \frac{\im}{2} (1\pm dw d\bar{w}).
\ee 

\subsection{Eigenstates}

We can now list eigenstates of the $\bK_3+\bP_3$ and $\bK_3-\bP_3$ operators. These operators only act in the space spanned by $z,\bar{z}, m,\bar{m}$ coordinates, which is a copy of $\R^4$. There are in total 16 differential forms in $\R^4$, 8 of each parity. These split into groups of four, those transforming non-trivially under $\bK_3+\bP_3$ and not transforming under $\bK_3-\bP_3$, and vice versa. 

Let us first list the states transforming non-trivially under ${\rm SU}(2)_R$, to which we take $\bK_3+\bP_3$ as belonging. We list the states in the format similar to that in the previous section. We get
\be
dz d\bar{w} \qquad \left( 0, \frac{1}{2}\right) \qquad \frac{1}{2}(1- dz d\bar{z})(1+dw d\bar{w}) \qquad \left( 0, -\frac{1}{2}\right), \\ \nonumber
\frac{1}{2}(1+ dz d\bar{z})(1-dw d\bar{w}) \qquad \left( 0, \frac{1}{2}\right) \qquad d\bar{z} dw \qquad \left( 0, -\frac{1}{2}\right) 
\ee
in the space of even forms in $\R^4$ and
\be
\frac{1}{\sqrt{2}} dz (1- dw d\bar{w}) \qquad \left( 0, \frac{1}{2}\right) \qquad \frac{1}{\sqrt{2}}  (1 -dz d\bar{z}) dw \qquad \left( 0, -\frac{1}{2}\right), \\ \nonumber
\frac{1}{\sqrt{2}} (1 +dz d\bar{z})d\bar{w} \qquad \left( 0, \frac{1}{2}\right) \qquad \frac{1}{\sqrt{2}} d\bar{z} (1+ dw d\bar{w})\qquad \left( 0, -\frac{1}{2}\right)
\ee
in the space of odd forms. As in the previous section, the complex conjugate of a right doublet is a right doublet, as it should be. 

The ${\rm SU}(2)_L$ doublets in the space of even forms in $\R^4$ are
\be
d\bar{z} d\bar{w}\qquad \left( \frac{1}{2}, 0\right) \qquad  \frac{1}{2}(1+ dz d\bar{z})(1+dw d\bar{w})  \qquad \left( -\frac{1}{2},0\right) ,\\ \nonumber
\frac{1}{2}(1- dz d\bar{z})(1-dw d\bar{w}) \qquad \left( \frac{1}{2},0\right) \qquad dz dw \qquad \left( -\frac{1}{2},0\right)
\ee
and in the space of odd forms
\be
\frac{1}{\sqrt{2}} d\bar{z} (1 -dw d\bar{w})  \qquad \left( \frac{1}{2},0\right) \qquad \frac{1}{\sqrt{2}} (1+ dz d\bar{z}) dw \qquad \left( -\frac{1}{2},0\right), \\ \nonumber
\frac{1}{\sqrt{2}}  (1- dz d\bar{z}) d\bar{w}\qquad \left( \frac{1}{2},0\right) \qquad \frac{1}{\sqrt{2}} dz (1 +dw d\bar{w}) \qquad \left(-\frac{1}{2},0\right).
\ee

The above are 16 forms in $\R^4$ and we need to complete them into even forms in $\R^7$. There are in total 8 forms in $\R^3$, 4 even and 4 odd. In each even/odd class two of the forms correspond to unprimed Lorentz spinor, and two to a primed one. We will only construct forms corresponding to unprimed spinors. The primed Lorentz spinors will then be obtained by complex conjugation. With this in mind, the forms in $\R^3$ corresponding to unprimed Lorentz spinors are $f^T_{even} \xi, f^T_{odd} \xi$, where $f_{even}, f_{odd}$ were introduced in (\ref{f-even}), (\ref{f-odd}). We now multiply the even forms in $\R^4$ by the even form $f^T_{even}\xi $, and odd forms in $\R^4$ by $f^T_{odd}\xi$ to get a convenient complex basis in the space of even forms in $\R^7$. It can then be checked that all the obtained states are also $\pm (\im/2)$ eigenstates of $T_{1,2,3}$, with the states that are ${\rm SU}(2)_R$ spinors having the property that the product of signs of eigenvalues is  always $-1$, and the product of sign for ${\rm SU}(2)_L$ spinors is $+1$. Then ${\rm SU}(4)$ acts by mixing the 4 left states, and the 4 right states. 

\subsection{The state $\Psi$}

We now introduce a convenient way to write the state $\Psi$ that is given by a general even form in $\R^7$, decomposed into the basis of eigenstates described above. First, the Lorentz spinor parts are described by forms $f_{even}, f_{odd}$. Let us introduce an analogous basis of states for ${\rm SU}(2)_{L,R}$ and ${\rm SU}(4)$. We will label these states by $f^{aI}$, where $a=1,2$ is an ${\rm SU}(2)$ index and $I=1,2,3,4$ is an ${\rm SU}(4)$ index. We view the states $f^{aI}$ with fixed $I$ as a 2-component row, and similarly for the state with fixed $a$, so that $f^{aI}\xi_{aI}$ makes sense. Then the right states are as follows
\be
f_R^{a1} = \left( \begin{array}{cc} dz d\bar{w} & \frac{1}{2}(1- dz d\bar{z})(1+dw d\bar{w}) \end{array}\right), \\ \nonumber
f_R^{a2} = \left( \begin{array}{cc} \frac{1}{2}(1+ dz d\bar{z})(1-dw d\bar{w}) & d\bar{z} dw \end{array}\right), \\ \nonumber
f_R^{a3} = \left( \begin{array}{cc} \frac{1}{\sqrt{2}} dz (1- dw d\bar{w}) &\frac{1}{\sqrt{2}}  (1 -dz d\bar{z}) dw \end{array}\right), \\ \nonumber
f_R^{a4} = \left( \begin{array}{cc}\frac{1}{\sqrt{2}} (1 +dz d\bar{z})d\bar{w} & \frac{1}{\sqrt{2}} d\bar{z} (1+ dw d\bar{w}) \end{array}\right),
\ee
and the left states are
\be
f_L^{a1} = \left( \begin{array}{cc} \frac{1}{\sqrt{2}} d\bar{z} (1- dw d\bar{w}) &\frac{1}{\sqrt{2}}  (1 +dz d\bar{z}) dw \end{array}\right), \\ \nonumber
f_L^{a2} = \left( \begin{array}{cc}\frac{1}{\sqrt{2}} (1 -dz d\bar{z})d\bar{w} & \frac{1}{\sqrt{2}} dz (1+ dw d\bar{w}) \end{array}\right), \\ \nonumber
f_L^{a3} = \left( \begin{array}{cc} d\bar{z} d\bar{w} & \frac{1}{2}(1+ dz d\bar{z})(1+dw d\bar{w}) \end{array}\right), \\ \nonumber
f_L^{a4} = \left( \begin{array}{cc} \frac{1}{2}(1- dz d\bar{z})(1-dw d\bar{w}) & dz dw \end{array}\right).
\ee
The reason why we changed the order and listed odd forms before the even forms for the left states will become clear below. 

We can now write the state $\Psi$ as
\be
\Psi = \Psi_L + \Psi_R + {\rm c.c.},
\ee
where
\be\label{psi-LR}
\Psi_L = f^T_{odd} f^{a1}_L \xi_{a1} + f^T_{odd} f^{a2}_L \xi_{a2} + f^T_{even} f^{a3}_L \xi_{a3} + f^T_{even} f^{a4}_L \xi_{a4}, \\ \nonumber
\Psi_R = f^T_{even} f^{a1}_R \bar{\xi}_{a1} + f^T_{even} f^{a2}_R \bar{\xi}_{a2} + f^T_{odd} f^{a3}_R \bar{\xi}_{a3} + f^T_{oddn} f^{a4}_R \bar{\xi}_{a4},
\ee
where every $\xi_{aI}, \bar{\xi}_{aI}$ is a 2-component unprimed Lorentz spinor represented by a column. Note that we could not have summed over the index $I$ in the above expressions because for right states $I=1,2$ are multiplied by $f_{even}$ while $I=3,4$ are multiplied by $f_{odd}$, and vice versa for the left states. 

\subsection{The hypercharge}

It is now an interesting exercise to work out the expression for the operator that gives the correct particles hypercharges, as is listed in e.g. Table J.1 of \cite{Dreiner:2008tw}. Some simple guesswork shows that the correct expression is
\be\label{Y}
Y = - \frac{1}{2} (\bK_3 + \bP_3) - \frac{1}{6} ( T_1+T_2+T_3).
\ee
Indeed, let us list the $- (1/6) ( T_1+T_2+T_3)$ charges of the states appearing in (\ref{psi-LR}). We have the following charges
\be\label{hyp-charges}
Y\left( \begin{array}{c} f^T_{odd} f^{a1}_L \\ f^T_{odd} f^{a2}_L \\ f^T_{even} f^{a3}_L \\ f^T_{even} f^{a4}_L \end{array} \right) = \left( \begin{array}{c} 1/6 \\ 1/6 \\ 1/6 \\ -1/2 \end{array} \right).
\ee
The operator $\bK_3+\bP_3$ does not act on the left states, and so this is already the correct assignment of hypercharges. This identifies the state with $I=4$ as lepton, while $I=1,2,3$ are the 3 different colours of quarks. 

For the right states the eigenvalues of $- (1/6) ( T_1+T_2+T_3)$ work out to be the opposite of those in (\ref{hyp-charges}), and adding to these the eigenvalues of $- (1/2)(\bK_3 + \bP_3)$ we get just the correct hypercharges for all the right states as well, as are listed in the Table J.1 of \cite{Dreiner:2008tw}. Again we have that $I=4$ describes the right-handed leptons $\bar{\nu}, \bar{e}$, while $I=1,2,3$ describes the 3 colours of the right-handed quarks $\bar{u}, \bar{d}$. 

The exercise that results in the expression for the hypercharge (\ref{Y}) is useful because it allows to identify which states correspond to quarks and which to leptons. It also shows how the choice of the hypercharge ${\rm U}(1)$ given by (\ref{Y}) explicitly breaks the ${\rm SU}(2)_R$ subgroup of the Pati-Salam group.

\subsection{The Lagrangian}

To compute the Lagrangian we need to compute the inner products of states introduced above. A calculation gives
\be
( (f_L^\dagger)_{a I}, f_L^{b J}) = (\im)^2 \delta_a^{b} \delta_I^{J} \epsilon(I),
\ee
where $\epsilon(I)=+1$ for $I=1,2$ and $\epsilon(I)=-1$ for $I=3,4$. For the right states the sign is opposite
\be
( (f_R^\dagger)_{a I}, f_R^{b J}) = -(\im)^2\delta_a^{b} \delta_I^{J} \epsilon(I).
\ee

This means that the part of $(\Psi, D\Psi)$ that depends on say $\xi_{a1}$ is
\be\label{a1}
(\xi^\dagger)^{a1} ( f^*_{odd} (f_L^\dagger)_{a1} , f^\dagger_{even} f^{b1}_L) \frac{1}{2} \partial \xi_{b1} = (\xi^\dagger)^{a1} ( f^*_{odd}  , f^\dagger_{even})(-1) ((f_L^\dagger)_{a1}, f^{b1}_L) \frac{1}{2} \partial \xi_{b1}  \\ \nonumber
(\xi^\dagger)^{a1}  (\im \epsilon) (\delta_a^b) \frac{1}{2} \partial \xi_{b1} = \im (\xi^\dagger)^{a1}\frac{1}{2} \epsilon\partial \xi_{a1}.
\ee
Here in the second step the extra sign appeared because we interchanged the odd degree forms. 
On the other hand, the part that depends on $\xi_{a3}$ is
\be
(\xi^\dagger)^{a3} ( f^*_{even} (f_L^\dagger)_{a3} , f^\dagger_{odd} f^{b3}_L) \frac{1}{2} \partial \xi_{b3} = (\xi^\dagger)^{a3} ( f^*_{even}  , f^\dagger_{odd})((f_L^\dagger)_{a3}, f^{b3}_L) \frac{1}{2} \partial \xi_{b3}  \\ \nonumber
(\xi^\dagger)^{a3}  (-\im \epsilon) (-\delta_a^b) \frac{1}{2} \partial \xi_{b3} = \im (\xi^\dagger)^{a3}\frac{1}{2} \epsilon\partial \xi_{a3}.
\ee
Thus, the final sign is the same as in (\ref{a1}). The final arising Lagrangian is
\be
(\Psi, D\Psi) = \frac{\im}{2} (\xi^\dagger)^{aI} \epsilon\partial \xi_{aI} - \frac{\im}{2} (\bar{\xi}^\dagger)^{aI} \epsilon\partial \bar{\xi}_{aI} + {\rm c.c.},
\ee
which is the result that has been quoted in (\ref{main-relation}) with the notation $\dirac:= \epsilon\partial$ used. Note that the ${\rm SU}(2)_R$ particles appear with the opposite sign in front of their kinetic term as compared to the left particles. This Lagrangian is manifestly ${\rm SO}(3,1)\times{\rm SU}(2)_L\times {\rm SU}(2)_R\times {\rm SU}(4)$ invariant as expected, and coincides with the kinetic term of the Standard Model fermion Lagrangian in which the lepton charge is treated as the fourth colour. If desired, it can be further split by separating $I=1,2,3$ states from $I=4$ states, which chooses an ${\rm SU}(3)$ subgroup of ${\rm SU}(4)$, and separates leptons from quarks. For more details on the used here 2-component spinor description of the SM we refer the reader to \cite{Dreiner:2008tw}, see Appendix J of this reference. 

\section{Discussion}

In this paper we carried out an exercise in ${\rm SO}(n,n)$ representation theory and worked out how the real Weyl spinor representation of this group splits under an ${\rm SO}(3,1)$ subgroup. We also dimensionally reduced the $\R^{n,n}$ Dirac operator to the Minkowski space $\R^{3,1}$. The outcome is a new interpretation of the fermion content of a single generation of the SM as a general real element of $\Lambda^{even}\R^7$, i.e. as a real inhomogeneous even degree differential form in $\R^7$. Thus, we have demonstrated that fermions, in particular fermions of relevance for the Standard Model, can be described by differential forms, in spite of the difficulties that are usually associated with this idea in the context of Dirac-K\"ahler fermions \cite{Kahler}. At the most basic level, the reason why everything works properly is that differential forms in half the dimension, i.e. in $\R^n$ rather than in $\R^{n,n}$, are used to describe spinors of ${\rm SO}(n,n)$. We also hope to have convinced the reader in the elegance of this formalism. In particular, the kinetic part of the SM fermion Lagrangian is correctly reproduced (\ref{main-relation}) by simple operations of exterior differentiation on $\R^{7,7}$ and applying the Clifford algebra relations (\ref{dx-clifford}). 

We believe this new interpretation is comparable in its elegance to something very well-known, namely the interpretation of the basic operators of vector calculus in $\R^3$ in terms of exterior differentiation of differential forms. In the vector calculus context, the 3 different differential operators turned out to be all just different incarnations of the same exterior derivative operator acting on different degree differential forms in $\R^3$. Something similar arises as the result of our construction. Indeed, different representations of the SM Pati-Salam gauge group ${\rm SU}(2)\times{\rm SU}(2)\times{\rm SU}(4)$ with their relevant Dirac operators are all seen to be different incarnations of a single even degree differential form in $\R^7$, with the Dirac operator being just an appropriately interpreted exterior derivative operator. The naturalness of our construction leads us to suggest that it is possibly the split signature group ${\rm SO}(7,7)$ that should be taken as the full group of "internal" symmetries of particle physics. 

Another interesting and simple model that we described is based on the ${\rm SO}(3,3)$ setup. In this case, the Weyl Lagrangian in $\R^{3,3}$ dimensionally reduces to a single charged Weyl fermion in the Minkowski space $\R^{3,1}$. This model may be a good playground for testing generalisations that are necessary to make the ${\rm SO}(7,7)$ model realistic. 

Even though we worked with just fermions, setting all the bosonic fields (apart from the metric) to zero, the construction that we described is clearly in the direction of unifying gravity with other interactions. This is clear from the fact that this construction puts together the Lorentz group, which is the group of local frame rotations that preserve the metric, with the gauge groups of the Standard Model. So, our spinor construction effectively unified the local gravitational symmetry with the usual gauge symmetry, by realising both as commuting subgroups of ${\rm SO}(7,7)$. 

A related remark is that the group ${\rm SO}(7,7)$ contains ${\rm GL}(7)$. This means that there is a link to the group of diffeomorphisms is seven dimensions. This is the reason why ${\rm SO}(7,7)$ transformations are related to the generalised diffeomorphisms in the double field theory context, see e.g. \cite{Hull:2014mxa} for a recent reference addressing this point. And if diffeomorphisms are symmetry, then gravity is present. 

The Lagrangian $(\Psi, D\Psi)$ that we obtained as the result of our construction only reproduces the kinetic part of the SM Lagrangian, with all bosonic fields (apart from the metric) set to zero. It is clear that the necessary step towards making the described in this paper ideas realistic is to understand how also the bosonic fields of the SM can be described. The most natural way to do this appears to be to make the global symmetry ${\rm SO}(7,7)$ of the Lagrangian $(\Psi, D\Psi)$ local, by extending the operator $D$ to some version of the covariant derivative. A mathematically natural way of doing this is to enlarge
\be\label{C}
D\to D+ C,
\ee
where $C$ is a general odd degree element of ${\rm Cliff}(T\oplus T^*)$. The reason why the restriction to odd degree elements arises is clear: The operator $D$ maps even forms to odd forms, and this is why the inner product in $(\Psi,D\Psi)$ gives a non-trivial result in odd dimensions. The extended covariant derivative operator should work in the same way, and hence $C$ must have odd degree. Further, it is probably not necessary to work with the most general odd degree element of ${\rm Cliff}(T\oplus T^*)$ because we act on Weyl spinors, and on these there is a relation between the action of an element of ${\rm Cliff}(T\oplus T^*)$ and the element that corresponds to its Hodge dual. Thus, it appears natural to restrict to say self-dual odd degree elements. The Lie algebra of ${\rm SO}(7,7)$, viewed as a subalgebra of ${\rm Cliff}(T\oplus T^*)$, then acts on the "connection" $C$ by the commutator. With this action the decomposition of the space of odd elements in ${\rm Cliff}(T\oplus T^*)$ into irreducibles under the Lorenz group ${\rm SO}(3,1)$ produces integer spins, which is as desired. Thus, the prolongation of the Dirac operator (\ref{C}) by an odd degree element of the Clifford algebra appears to be the most mathematically natural way of coupling our fermions to bosons. It remains to be seen if these ideas can be made realistic. 

Let us also remark that it is the fact that $C$ in (\ref{C}) must be of odd degree that made us choose to describe fermions by even degree forms in the first place. Thus, the proposal is that fermions are to be described by forms in $\Lambda^{even}\R^7$, while bosons $C$ should be described by $\Lambda^{odd} \R^{7,7}$ (possibly self-dual elements thereof, and possibly with some additional restrictions imposed), where we used the identification of the Clifford algebra with the exterior algebra. 

If one is to take the described geometric construction seriously, the main open question is what mechanism breaks the ${\rm SO}(7,7)$ symmetry down to what is seen in Nature. We do not yet have any satisfactory answer to this question, but there are some hints indicating that it may be possible and natural for such a mechanism to exist, as we now describe. 

We first remark that ${\rm SO}(7,7)$ is special from all ${\rm SO}(n,n)$ groups, being the group of largest dimension where a Weyl spinor has a non-trivial stabiliser so that (a dense subset of) the space of Weyl spinors can be viewed as an $\R^*\times {\rm SO}(n,n)$ orbit, where $\R^*$ acts by rescaling. The fact that the space of spinors can be given the interpretation of a group coset plays the crucial role in the generalised geometry setup \cite{Hitchin:2004ut}, where it is ${\rm SO}(6,6)$ that is relevant. In the ${\rm SO}(7,7)$ case studied in \cite{Witt:2005sk} a unit Weyl spinor is shown to have the stabiliser $G_2\times G_2$ or $G_2(\C)$, where in the former case both real forms of the exceptional Lie group $G_2$ can appear. The largest dimension when this phenomenon occurs is $n=7$, as a simple comparison of dimensions of relevant spaces shows. We find the coincidence between the dimension of the largest ${\rm SO}(n,n)$ that admits an orbit interpretation of the space of spinors, and the dimension of ${\rm SO}(n,n)$ that appears from the SM fermions interpretation striking. 

This suggests that the group ${\rm SO}(7,7)$ can be broken by a fixed Weyl spinor. As we have said, the possible unbroken subgroups in this case are $G_2\times G_2$ or $G_2(\C)$. The last of these contains the Lorentz group ${\rm SL}(2,\C)$. Moreover, because $G_2\times G_2\subset {\rm SO}(7)\times {\rm SO}(7)$ for the compact $G_2$ (for the non-compact real form of $G_2$ this relation is $G_2\times G_2\subset {\rm SO}(4,3)\times {\rm SO}(4,3)$) and $G_2(\C)\subset {\rm SO}(7,\C)$, in all cases the stabiliser group defines a split $V=U\oplus \tilde{U}$ and the metric $G$ restricts to non-degenerate metrics in $U,\tilde{U}$, or a complex metric in $U$ in the case of $G_2(\C)$ stabiliser. This naturally brings the metric geometry into play, something exceptional that only happens in the ${\rm SO}(7,7)$ setup. Thus, after the ${\rm SO}(7,7)$ symmetry is broken by the split $V=U\oplus\tilde{U}$, gravity automatically arises as a part of the present geometric construction. 

A related way of looking at the ${\rm SO}(7,7)$ symmetry breaking is the already discussed necessity of choosing a Lorentz subgroup ${\rm SO}(3,1)\subset{\rm SO}(7,7)$. There are many ways of doing this, and it is in the process of making this choice that the metric (=gravity) appears, and also 4 out of 7 coordinates in $\R^7$ are chosen as special. However, to make all these ideas concrete, one has to make the mechanism by which the Lorentz group is selected dynamical. This is only possible if we have a dynamical principle for the bosonic fields $C$ in (\ref{C}). Presumably, it should be of the first order type $CDC$, plus a potential for $C$, but details are still to be worked out. 

There is yet another geometric construction that shows that gravity and metric interpretation arises naturally in the context of odd degree differential forms in $\R^7$ is that in \cite{Krasnov:2016wvc}, \cite{Krasnov:2017uam}. In this references it was shown that (Euclidean signature) gravity in four dimensions can be understood as the dimensional reduction of a theory of differential 3-forms in seven dimensions. The metric interpretation arises because of the classical fact that a generic 3-form in seven dimensions defines a metric, see \cite{Krasnov:2016wvc}, \cite{Krasnov:2017uam} for more details. 

All in all, there are several not obviously related geometric constructions that all suggest that there is a natural metric interpretation of the  ${\rm SO}(7,7)$ setup, and this interpretation is only possible for $n=7$, which is also the largest dimension where the group ${\rm SO}(7,7)$ has certain desirable properties. There also appears to be a natural mechanism in place to break the ${\rm SO}(7,7)$ symmetry, and the metric interpretation becomes possible after this symmetry is broken. If one adds to this list the geometric construction we described in this paper, which shows that also the SM fermions have an ${\rm SO}(7,7)$ interpretation, one is forced to conclude that all roads lead to ${\rm SO}(7,7)$.

\end{document}